\documentstyle[floats,psfig,aps,prd,amssymb]
{revtex}

\def\he#1{$^{#1}$He}
\def\li#1{$^{#1}$Li}

\def\h#1{$^{#1}$H}

\def\pbar{$\bar p$}
\def\mpbar{\bar{p}}

\def\mhe#1{{^{#1}\mbox{He}}}
\def\mli#1{{^{#1}\mbox{Li}}}

\newcommand{\pr}{Phys. Rev. }
\newcommand{\aap}{Astron. Astrophys.}
\newcommand{\mnras}{Mon. Not. Roy. Astron. Soc. }

\newcommand{\apss}{Astrophys. Sp. Science }

\newcommand{\bref}[1]{(\ref{#1})}               
\newcommand{\ra}{r_{\rm A}^{100}}
\newcommand{\nbb}{n_{\bar b}}
\newcommand{\np}{n_\gamma}
\newcommand{\net}{n^{\rm net}} 
\newcommand{\bnet}{\bar{n}^{\rm net}} 
\newcommand{\neta}{\eta_{\rm net}}

\newcommand{\tot}{{\rm d}}
\renewcommand{\phi}{\varphi}
\renewcommand{\rho}{\varrho}
\newcommand{\nbar}{$\bar{n}$}

\newcommand{\cm}{\,{\rm cm}}
\newcommand{\s}{\,{\rm s}}

\newcommand{\mb}{\,{\rm mb}}
\newcommand{\MeV}{\,{\rm MeV}}
\newcommand{\GeV}{\,{\rm GeV}}
\newcommand{\keV}{\,{\rm keV}}

\newcommand{\RA}{{R_{\rm A}}}
\newcommand{\Tanni}{{T_{\rm ann}} }
\newcommand{\svanni}{{ \sigma_{\rm ann}v_b} } 
\newcommand{\fb}{f_{\bar b}}
\newcommand{\The}{{T_{^4{\rm He}}}}
\newcommand{\simge}{\gtrsim}
\newcommand{\simle}{\lesssim}

\begin{document}

\draft
\twocolumn[\hsize\textwidth\columnwidth\hsize\csname
@twocolumnfalse\endcsname
\title{Limits on Cosmic Matter--Antimatter Domains
from Big Bang Nucleosynthesis}

\author{Jan B. Rehm\thanks{Electronic address: jan@mpa-garching.mpg.de} 
and Karsten Jedamzik\thanks{Electronic address: jedamzik@mpa-garching.mpg.de} 
}
\address{
Max-Planck-Institut f\"ur Astrophysik, Karl-Schwarzschild-Str. 1,
85748 Garching, Germany
}

\date{\today}
\maketitle

\begin{abstract}
We present detailed numerical calculations of the light element
abundances synthesized in a Universe consisting of matter- and
antimatter- domains, as predicted to arise in some electroweak
baryogenesis scenarios. In our simulations all relevant physical
effects, such as baryon-antibaryon annihilations, production of
secondary particles during  annihilations, baryon diffusion, and
hydrodynamic processes are coupled to the nuclear reaction network.
We identify two dominant effects, according to the typical spatial
dimensions of the domains.  Small antimatter domains are dissipated
via neutron diffusion prior to \he4 synthesis at $\The \approx
80$~keV, leading to a suppression of the primordial \he4 mass
fraction. Larger domains are dissipated below $\The$ via a combination of
proton diffusion and hydrodynamic expansion.  In this case the strongest
effects on the elemental abundances are due to \pbar\he4
annihilations, leading to an overproduction of \he3 relative to
\h2 and to overproduction of \li6 via non-thermal nuclear reactions.
Both effects may result in light element abundances deviating
substantially from the standard Big Bang Nucleosynthesis yields and
from the observationally inferred values. This allows us to derive
stringent constraints on the antimatter parameters. For some
combinations of the parameters, one may obtain both, low \h2 and
low \he4, at a common value of the cosmic baryon density, a result
seemingly favored by current observational data.

\end{abstract}

\pacs{26.35.+c,98.80Cq,25.43.+t}
]
\narrowtext

\section{Introduction}\label{S:Intro}
\footnotetext{$^\ast$ Electronic address: jan@mpa-garching.mpg.de} 
\footnotetext{$^\dagger$ Electronic address: jedamzik@mpa-garching.mpg.de}
Recently, there has been a revived interest in antimatter cosmologies,
stimulated in part by the first flight of the Alpha Magnetic
Spectrometer (AMS) and by the prospect of a long-term AMS mission on
board of the International Space Station Alpha (ISSA). While these
efforts concentrate on direct detection of antinuclei in the solar system,
several limits on the existence of antimatter domains 
have been placed in the
past. It has been known for a long time, that the presence of
significant amounts of antimatter within a distance of about 20 Mpc
from the solar system may be excluded on grounds of the
non-observation of annihilation radiation~\cite{St:76}. More recent
studies of the diffuse $\gamma$-ray background claim to exclude
antimatter regions in today's Universe within a distance of $\sim$ 1000 Mpc
\cite{CRG:98}, a considerable fraction of the present horizon
$\sim 3000\,$Mpc. The existence of antimatter domains would also have impact
on the Cosmic Microwave Background Radiation (CMBR).
Recent studies predict
`ribbon'- or `scar'-like anisotropies in the CMBR, at the interfaces
of matter- and antimatter domains, 
with a Sunyaev-Zel'dovich-type $y$ distortion of the order of
$10^{-6}$~\cite{KKT:97,CR:98}.  
Such small distortions are beyond the detection limits of
current CMBR observations, and probably also beyond those of
the upcoming MAP and PLANCK satellite missions.
Given that only a small region of
the parameter space for the sizes of antimatter domains
remains, it seems very unlikely that
we live in a Universe containing any considerable amount of antimatter
today. In particular, the possibility of a baryo-symmetric Universe
containing equal amounts of matter and antimatter is excluded unless
the separation length of matter and antimatter is nearly as large as the
current horizon. It is, however, not possible on grounds of the above
results to exclude the existence of small and distant pockets of
antimatter~\cite{DS:93}.

A complementary scenario with small scale domains of
antimatter which have completely 
annihilated prior to recombination has, however, hardly been
investigated in the past.
Note that such a scenario necessarily involves an excess
of matter over antimatter.
While the very precise observation of the CMBR allows us to place
stringent limits
on any non-thermal energy input into the CMBR during
epochs with CMBR temperature $0.3\, {\rm eV}\lesssim$ $T\lesssim 1\,$ keV,
in particular, also on energy input due to annihilations~\cite{SZ:70b},
even more stringent constraints may be derived from considerations of
Big Bang Nucleosynthesis (BBN, hereafter). 
Such a baryo-asymmetric Universe filled with a distribution of small-scale
regions of matter or antimatter may, for example, 
arise during an epoch of baryogenesis
at the electroweak scale. It has been shown within the minimal
supersymmetric standard model, and under the assumption of explicit as
well as spontaneous $CP$ violation, that during a first-order
electroweak phase transition the baryogenesis process may result in
individual bubbles containing either net baryon number, or net
anti-baryon number\cite{CPR:94}.  More recently, it has been argued
that pre-existing stochastic (hyper)magnetic fields in the early
Universe, in conjunction with an era of electroweak baryogenesis, may
also cause the production of regions containing either matter or antimatter
\cite{GS:98ab}. Though it seems questionable at present
if electroweak baryogenesis has occurred at all,
there are other imaginable scenarios which could lead
to a small-scale matter-antimatter domain structure in the early
Universe~\cite{KRS:00}.

This kind of initial conditions may have profound consequences on the
abundances of the cosmological synthesized light elements.  In the
standard picture, synthesis of the light elements
takes place between the cosmological epoch
of weak freeze out at $T\approx 1 $~MeV and $T\approx 20
$~keV. The abundances of the light elements are highly sensitive to
the cosmic conditions during that epoch. For example,
the primordially synthesized \he4 mass fraction is sensitively dependent 
on the relative
abundances of protons and neutrons at $\The \approx 80$~keV, when
practically all available neutrons are incorporated into \he4 nuclei.
We have recently shown
that annihilation of antimatter domains during BBN (at temperatures above 
$\The$) may significantly alter the neutron-to-proton
ratio at $\The$~\cite{RJ:98}. However, light
element abundances are also sensitive to putative matter-antimatter
annihilations after the epoch of BBN.
In this paper, we thus extend
our analysis of annihilation of antimatter domains to a much wider
temperature regime, from above the epoch of weak freeze out
to the epoch of recombination. This allows us to constrain 
matter-antimatter domains within
a much wider range of domain separations. 
The same scenario, annihilation of matter-antimatter domains during
and after BBN, has been very recently investigated in a Letter by 
Kurki-Suonio and Shivola.~\cite{KS:99}. Though the main conclusions of
our paper are not vastly different from those of Ref.~\cite{KS:99},
we arrive at somewhat different results (factor $\sim 3$) 
for the synthesis of some of the elemental
abundances. Furthermore, we
use a different approach in comparing our theoretical results with
observationally inferred values for the light element abundances. In
particular, we base our constraints on observationally determined
limits on the primordial $^3$He/\h2 ratio, rather than on the much less secure
limit on primordial $^3$He. We also consider the production of $^6$Li
which yields a tentatively much more stringent limit on the existence
of antimatter domains.  
 
Prior studies of the influence of antimatter domains on the light element
abundances  have only been carried out  in
the context of a baryo-symmetric Universe
\cite{St:76,St:72_CFL:75_Al:78}. Of course, such models have to assume
that annihilation of all cosmic baryons may be avoided by an assumed
\lq\lq unphysical\rq\rq\ , and unknown rapid separation mechanism of
matter from antimatter.
In essence, these works have shown 
that antimatter domains and successful BBN mutually exclude each other in
baryo-symmetric cosmologies unless the separation between matter and 
antimatter domains is exceedingly large. In that case, however, BBN
proceeds in a  
standard way, independently in matter and antimatter domains.

There have been a number of studies concerning a {\it homogeneous}
injection of antimatter into the primordial plasma during, or after 
BBN~\cite{CKSZ:82_CKS:82,BFPS:84,KL:84_ENS:85_Li:86_Do:87_Ha:87,DY:87_YD:88}.
Antibaryon production may result through the decay of relic, heavy
particles $X$, if hadronic decay channels are present, or the
evaporation of primordial black holes.  A possible candidate for the
$X$-particle is the gravitino, the superpartner of the graviton.  It
was realized early that antibaryons injected around weak freeze out
would increase the synthesized $^4$He abundance, due to preferential
annihilation on protons~\cite{ZSKC:77_VDN:78}. This was used to derive
a limit on the relative antibaryon abundance of $\nbb/(n_n+n_p)
\lesssim 1/20$. Injection of antibaryons after BBN would result in
\pbar\he4 annihilations~\cite{CKSZ:82_CKS:82}. The concomitant
production of $^2$H, $^3$H, and $^3$He, either by direct production
during $^4$He annihilation, or by fusion processes of secondary
neutrons, was found to give unacceptable large ($^2$H + $^3$He)/H,
unless $\nbb/n_b \lesssim 10^{-3}$~\cite{BFPS:84}.  Such arguments were
subsequently used to constrain the abundances of particular relic
particles~\cite{KL:84_ENS:85_Li:86_Do:87_Ha:87}. 
Motivated by the idea to reconcile a Universe
dominated by baryonic dark matter with BBN, Yepes \&
Dom\'{\i}nguez-Tenreiro investigated the effects of antibaryons
injected {\it during} BBN~\cite{DY:87_YD:88}. An outstanding feature of such
scenarios is a significant reduction of $^7$Li production compared to
standard BBN. Nevertheless, the claim that such scenarios may be
compatible with a fractional contribution of baryons to the critical
density of $\Omega_b = 1$ seems not viable due to the overproduction
of \he4 and the $^3$He/$^2$H ratio.

In this paper we investigate BBN with matter-antimatter domains.
Assumed initial conditions and definitions which are used throughout
the paper are introduced in Sec~\ref{S:Pre}.
In contrast to homogeneous antimatter injection, this topic
requires an understanding of 
hydrodynamic and diffusive processes which lead to
the mixing of matter and antimatter. 
These processes are summarized in Sec.~\ref{S:hydro}. 
Nuclear annihilation reactions and the production of secondaries and
their evolution are investigated in Sec.~\ref{S:nucl}. In
Sec.~\ref{S:bbnma} we then present results of detailed numerical
simulations of BBN in presence of matter-antimatter domains.
New and stringent limits on antimatter domains are derived in
Sec.~\ref{S:Res}, whereas Sec.~\ref{S:Disc} is devoted to discussion and
conclusions. The appendices discuss the structure of the actual 
annihilation region (App.~\ref{S:ann_regio}) and some aspects of the numerical
treatment of the problem (App.~\ref{S:num}).
\begin{figure}
   \centerline{\psfig{figure=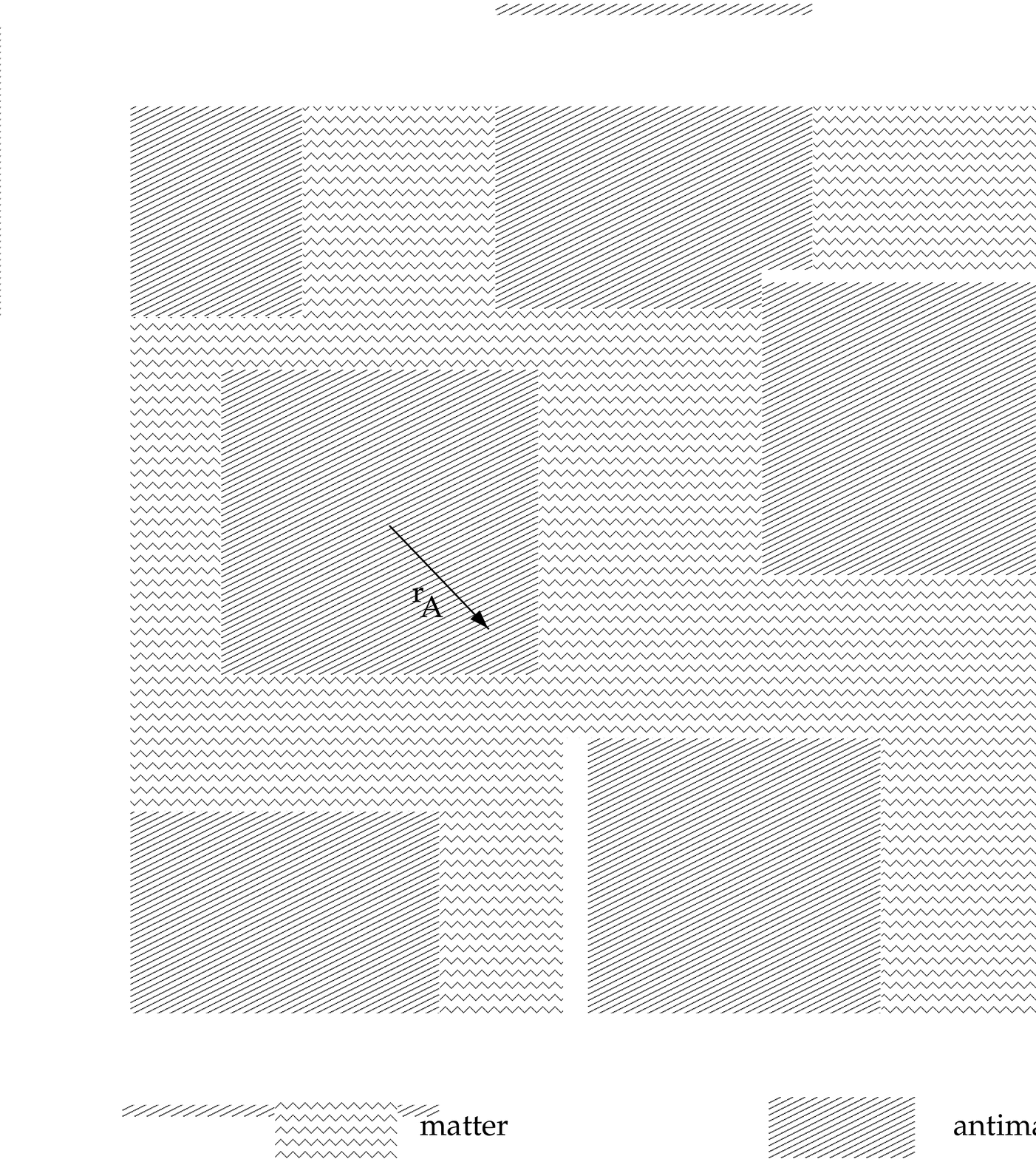,width=3.375in}}
{\caption{Sketch of the assumed initial conditions: Antimatter domains 
are embedded in a background of matter.
  \label{F:sketch}}}
\end{figure}
\section{Preliminaries}\label{S:Pre}
In this work, we consider cosmological models which contain different
amounts of antimatter distributed in domains of various sizes 
$r_{\rm A}$.
In such scenarios, the Universe may be envisioned
as a distribution of
matter with embedded domains of antimatter, as is schematically shown
in Fig.~\ref{F:sketch}.  Initially, the matter densities
$n_b(r|r>r_{\rm A})$ in the matter region and antimatter densities $n_{\bar
b}(r|r<r_{\rm A})$ in the antimatter domains with radius $r_{\rm A}$
are assumed to be equal throughout the Universe.  The average net
baryon density is thus given by
\begin{equation}
\bnet \equiv \bar\neta\bar\np= n_b(r|r>r_{\rm A})(1-\fb)
-\nbb(r|r<r_{\rm A})  \fb\;,\label{E:net} 
\end{equation}
where $\bar{n}_\gamma$ is the average photon number density,
$\bar{\eta}_{\rm net}$ the average net baryon-to-photon ratio, and the
filling factor $\fb$ is defined as the fraction of the volume of the
Universe filled with antimatter.  Here and in the following a bar over
some quantity denotes the horizon average of that quantity. Since
baryo-symmetric models will not leave baryon number after the
completion of annihilation, models with an excess of baryon number
will be considered, i.e. $\fb < 0.5$. Further, we define the
antimatter-to-matter ratio
\begin{eqnarray}
\RA\equiv N_{\bar b}/N_b
\end{eqnarray} 
as the ratio of
antibaryon number $N_{\bar b} = \int_V \nbb(r) \tot^3r$ to baryon number 
$N_b = \int_V n_b(r) \tot^3r$.

It is convenient to express the length scales in our problem in
comoving units. The length of, e.g., an antimatter region 
at some cosmic time $t$, or equivalently temperature  $T$, may be related
to  the length
it had at a fixed temperature $T_0$, which we choose to be 100~GeV.
The physical size $l(T)$ of that region in terms of the comoving size
$l_{100} \equiv l(100 \GeV)$ is thus given by
\begin{eqnarray}
l(T) = l_{100} \, \left(\frac{R(T)}{R_{100}}\right)\; ,
\label{E:lR}
\end{eqnarray} 
where $R(T)$ is the cosmic scale factor at an epoch with temperature
$T$, and we define $R_{100}\equiv R(100 \GeV) =1$.  The time evolution
of the scale factor may be derived from the conservation of entropy,
$S\propto g_{\ast s} T^{3}R^3 = \mathit{const.}$ Thus the scale factor
evolves as $R \propto g_{\ast s}^{-1/3} T^{-1}$, where $g_{\ast s}$ is
the number of relativistic degrees of freedom contributing to the
entropy density of the Universe. At the electroweak phase transition
($T\approx 100 \GeV$), $g_{\ast s} \approx 100$. For
definiteness, we will assume that $g_{\ast s}(100 \GeV) = 100$, which
allows us to calculate the physical length $l$ as a
function of the comoving length at 100 GeV, $l_{100}$. The comoving
scale $l_{100}$ thus corresponds, for example, to a physical length at
a temperature of 1~MeV of $l(1 \MeV) = l_{100} \times 10^5
(100/10.75)^{1/3}$, and to $l(T) = l_{100}\, (100 {\rm GeV}/T)
(100/3.909)^{1/3}$, for any epoch subsequent to
$e^{\pm}$-annihilation.

The time evolution of the densities $n_i(r,t)$ of the 
nucleon or light nuclei species $i$
(and their antiparticles)
is governed  by three
mechanisms, namely diffusion and hydrodynamic processes, annihilation,
and nuclear reactions, 
\begin{eqnarray}
\frac{\partial n_i}{\partial t}= 
\left.\frac{\partial n_i}{\partial t}\right|_{\rm diff/hydro}
+\left.\frac{\partial n_i}{\partial t}\right|_{\rm ann}
+\left.\frac{\partial n_i}{\partial t}\right|_{\rm nuc}.
\label{E:nuc_evol_full}
\end{eqnarray}
We discuss diffusive and hydrodynamic processes in 
Sec.~\ref{S:hydro}, whereas baryon-antibaryon
annihilations will be discussed in Sec.~\ref{S:nucl}. The
nuclear reaction network has been widely discussed in the literature
(e.g.~\cite{WFH:67,SKM:93}) and will thus not be described
here. Some aspects of the numerical treatment may be found in
App.~\ref{S:num}.

We will frequently use variables 
$\Delta_i(r)$, which are defined in terms of the number 
density of species $i$
at coordinate $r$ and the average net baryon number density
$\bar\net$
\begin{eqnarray}
\Delta_i(r) = \frac{n_i(r)}{\bar\net}\; .
\label{E:bary_overdens}
\end{eqnarray}
We also will use the quantity $\delta (r)$ which denotes the relative
cosmic temperature variation at $r$, i.e.
\begin{equation}
\delta (r) = \frac{T(r)-\bar{T}}{\bar{T}}\; ,
\label{E:delta}
\end{equation} 
where $\bar{T}$ is an appropriately defined cosmic average temperature
and $T(r)$ is the local temperature.
\section{Diffusive and Hydrodynamic Processes}\label{S:hydro}
\subsection{Pressure Equilibrium}
\begin{figure}[t]
   \centerline{\psfig{figure=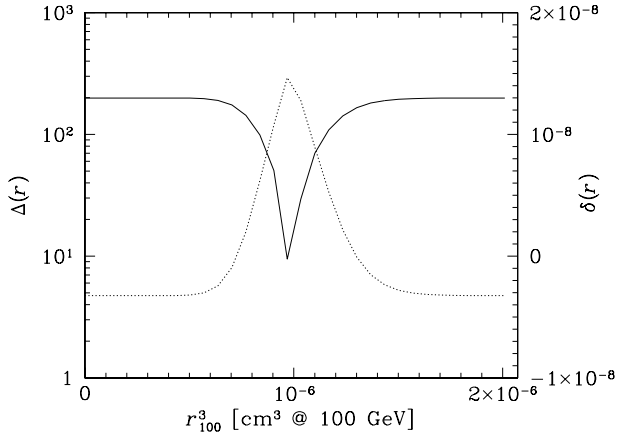,width=3.375in}}
{\caption{Snapshot of an isobaric density fluctuation. The full 
        line represents the total baryonic (i.e. baryonic- plus 
        antibaryonic-) overdensity
        $\Delta(r)$, the dotted line the deviation from the average
        cosmic temperature $\delta(r)$. The boundary between the matter-
        and the antimatter- domain is approximately at
	$r_{100}^3\approx 10^{-6}$,  
        where $r_{100}$ is the radius of the spherical simulation volume. 
\label{F:isobaric}}}
\end{figure}

Our initial conditions are such that
matter and antimatter regions exist in pressure equilibrium with
each other at uniform cosmic temperature. 
As long as the transport of
baryon number over the boundaries from one region into the other is
not efficient, matter and antimatter are kept in separate regions.  The
photon and lepton densities are homogeneous and the temperature is the
same throughout the Universe.  Inhomogeneities in the total baryonic
density, which is defined as the baryon- plus antibaryon- density at
position $r$, $|n_b^{\rm tot}(r)|=n_b(r) +\nbb(r)$, may arise only
when transport of baryon number over the domain boundaries occurs
and annihilation proceeds.  Subsequently, the baryon and antibaryon
densities close to the boundary decrease, leading to a decrease in the
(anti-) baryon pressure in the annihilation region.  This baryonic
underpressure is then compensated for by a slight adiabatic
compression of the region and thus an increase of the radiation
pressure, as to reestablish pressure equilibrium between the region
and its environment.  
Fluctuations which have come into pressure
equilibrium will be termed `isobaric' fluctuations~\cite{JF:94}.
Note that the time scale for reestablishing pressure equilibrium is
by far the shortest of all time scales in our problem, such that the
assumption of attaining pressure equilibrium instantaneously is justified.
An isobaric fluctuation after some annihilation at the domain
boundaries has occurred is shown in Fig.~\ref{F:isobaric}.

At late times, and for fluctuations where the photon mean free path
after $e^\pm$ annihilation becomes large compared to the spatial scale
of the fluctuations (typically at $T\approx 20\,$ {\rm keV}), 
temperature gradients between the fluctuations
cannot be maintained any more and the assumption of pressure
equilibrium breaks down.  
In this regime the density inhomogeneities are dissipated by
hydrodynamic expansion of the cosmic fluid (see below).
\subsection{Baryon Diffusion}
Transport of (anti-) baryon species $i$ 
over the domain boundaries may only be
accomplished by diffusion, which is described by
\begin{eqnarray}
\frac{\partial n_i}{\partial t} = D_i \nabla^2n_i\; , \label{E:diff}
\end{eqnarray}
where $D_i$ is the relevant diffusion constant.
Equation~\bref{E:diff} may  be written in comoving radial coordinates
(see Ref.~\cite{JF:94}) 
\begin{equation}
\frac{\partial \Delta_i(T,r_{100})}{\partial
t}=\frac{1}{r_{100}^2}\frac{\partial 
}{\partial r_{100} }\left(\frac{D_i}{R^2} r_{100}^2 \frac{\partial
}{\partial r_{100}} 
\Delta_i(T,r_{100})\right)\; .
\end{equation}
where we have used the notation introduced in the previous section.

The diffusion constant $D_{ik}$ for baryon species $i$ due to
scattering off some species $k$ with cross section $\sigma_{ik}$ and
number density $n_k$ is approximately given by the product of thermal
baryon velocity $v_i$ and baryon mean free path $l_{ik}$ of the particle
under consideration,
\begin{eqnarray}
D_{ik}\approx \frac{1}{ 3}v_il_{ik}=
\frac{1}{3}v_i\frac{1}{\sigma_{ik}n_{k}}\; .
\end{eqnarray}
Some relevant diffusion constants and their cosmological importance
may be found in Ref.~\cite{JF:94}. The effective baryon diffusion constant
of nucleus $i$ in the plasma due to  scattering off different species  $k$
is given by 
\begin{eqnarray}
\frac{1}{D_i}=\sum_k \frac{1}{D_{ik}}\; .
\end{eqnarray}
The diffusion length of a species is defined as the rms distance
traveled during time $t$. Written in comoving coordinates, one finds
(see Ref.~\cite{AHS:87,JF:94})
\begin{eqnarray}
d_{100}(t)=\left[6\int_0^t R^{-2} D(t')\tot t'\right]^{1/2}.\label{E:diff_len}
\end{eqnarray}
Three different temperature regimes with respect to the
diffusion of baryons may be distinguished. At early times, prior to
the annihilation of thermal 
electron-positron pairs, the proton diffusion length is short due to
their electromagnetic interaction with the ambient pairs. Neutron
diffusion is controlled by the much weaker magnetic moment
scattering on electron-positron pairs and thus the diffusion length is
longer.  As long as the
temperature is higher than $\approx 1$~MeV, neutrons and protons are
however constantly interconverted by the fast weak
interactions. During this epoch, 
baryons may thus diffuse during the time they spend as
neutrons~\cite{AHS:87}.
After
the weak interactions freeze out at $T\approx 1$~MeV, neutron-to-proton
interconversion ends, but the neutrons continue to diffuse. At $\The \approx
80$~keV, virtually all free neutrons are bound into \he4 nuclei. All
baryons and antibaryons exist now in the form of charged nuclei and
antinuclei.

Proton and charged nuclei diffusion is limited by Coulomb scattering 
off electrons and
positrons from the time of weak freeze out down to a temperature
$T\approx 40$~keV.
The Coulomb cross section for the light nuclei is proportional to
the square of the nuclear charge $Z_i^2$  and
the thermal velocity to $\sqrt{1/A_i}$, where $A_i$ is the mass number of the
nucleus under consideration. This leads to a suppression factor of
the diffusivity of nuclei relative to that for protons of
$(1/Z_i^2\sqrt{A_i})$~\cite{remark}. 

\begin{figure}
   \centerline{\psfig{figure=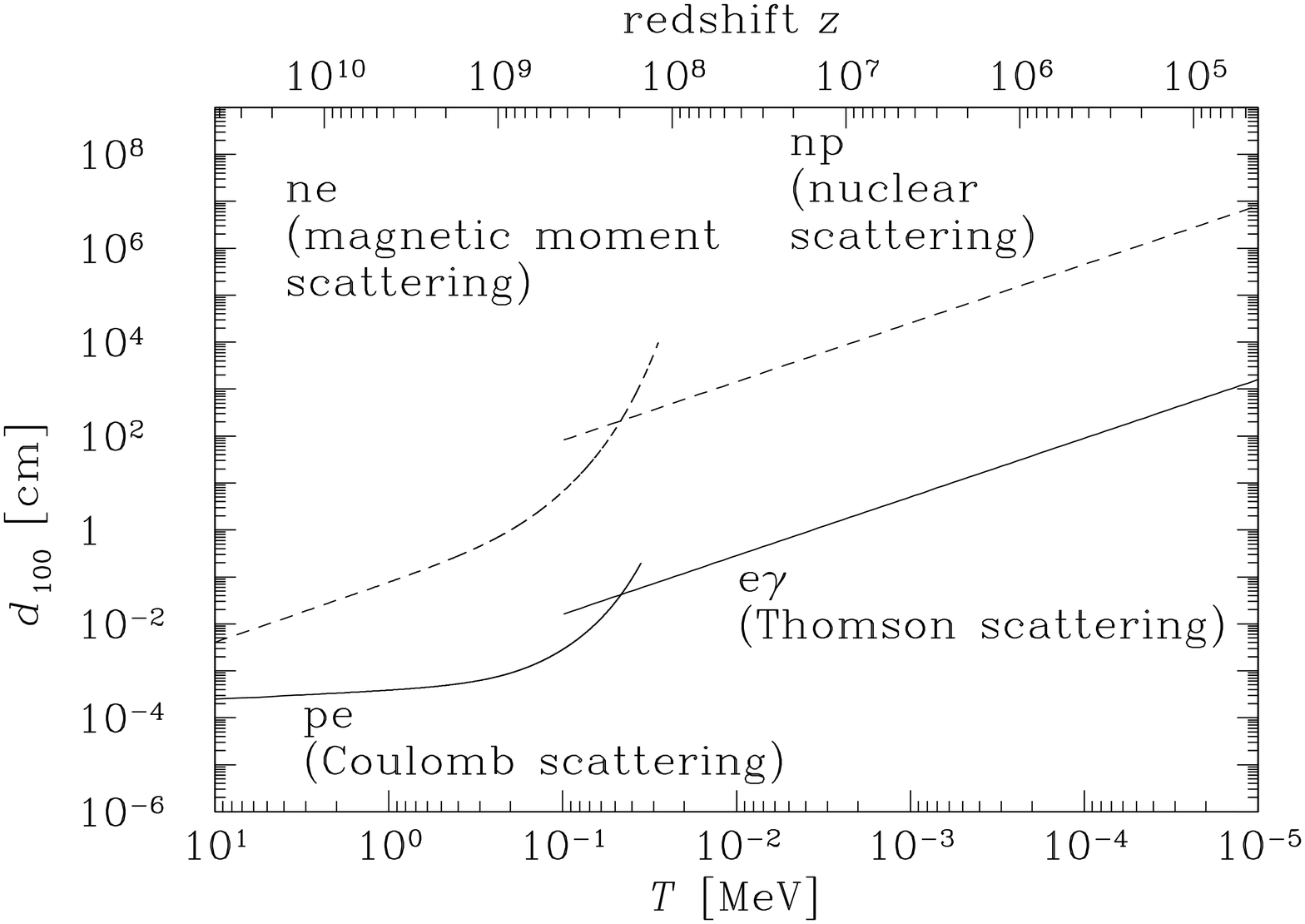,width=3.375in}}
{\caption{Diffusion lengths $d_{100}$ of (anti-) neutrons (dashed lines) and
        (anti-) protons (full lines) as a function of cosmic temperature, 
        measured on the 
	comoving scale fixed at $T=100 \GeV$. Here 
        a baryon-to-photon ratio of $\eta = 4\times
        10^{-10}$ has been assumed.
  \label{F:diff}}}
\end{figure}

When the  pair density decreases at temperatures lower than 
$T\approx 40$~keV, protons cease to diffuse as individual particles. Rather, a
proton-electron system diffuses together in order 
to maintain electric charge neutrality and consequently the larger
electron photon cross section dominates the proton diffusion 
constant~\cite{JF:94}. Diffusivity of nuclei at these temperatures is
expected to be suppressed by a factor $1/Z_i$ relative to that of protons.
The proton and electron diffusion lengths are displayed in
Fig.~\ref{F:diff} for the temperature range of interest.  
\subsection{Heat Diffusion and Hydrodynamic Expansion}
Heat transport between isobaric high and low density regions, in particular
between regions with high and low total baryonic pressure and concomitant low
and high radiation pressure, may be
accomplished by diffusing or free streaming neutrinos or by diffusing
photons.
Note that the effect of such heat transport is the decrease of (anti-)
baryonic density in overdense regions, and the increase of (anti-)
baryonic density in underdense regions.
We will be interested in the
evolution of inhomogeneities which are generated 
by annihilations at relatively
low temperatures $T \lesssim {\it a\; few}$~MeV, such that neutrino heat
conduction is typically inefficient~\cite{JF:94}.
In contrast, heat transport via photons 
may be efficient towards the end of the 
$e^\pm$ annihilation, depending on the length scale 
of the temperature fluctuations. 
During the time of $e^\pm$ annihilation, the comoving photon mean free
path
\begin{eqnarray}
l^{\gamma e}_{100} \approx \frac{R^{-1}}{\sigma_{\rm T} n_{e^\pm}}
\end{eqnarray}
increases enormously since it is inversely proportional to the total
number density of electrons and positrons,
$n_{e^\pm}=n_{e^+}+n_{e^-}$. At 
temperatures below $T\,\simle\, 30\,$keV essentially all pairs have 
annihilated and $n_{e^\pm}$ is dominated by net electron number
densities required to maintain charge neutrality. 
The increased photon mean free path may then affect
the dissipation of fluctuations in the baryon
density~\cite{ADFMM:90}. 
As long as the photon mean free path is still shorter than the scale
of the fluctuation, heat transport is described by the diffusion equation for
photons, which is identical to Eq.~\bref{E:diff}, but with $\Delta_i(r)$
replaced by the temperature fluctuations $\delta(r)$. The diffusion
constant is now given approximately by
\begin{eqnarray}
D_\gamma\approx\frac{g_{t}}{g_\ast} l_{\gamma e}
\end{eqnarray}
with $g_{t}$ the statistical weight of the heat transporting
particles ($g_{t}=2$ for photons) and  $g_\ast$ the
statistical weight of the relativistic particles still coupled to the plasma
($g_\ast= g_{t}$ after $e^\pm$ annihilation, since neutrinos
are decoupled).

When the photon mean free path becomes larger than the scale
of fluctuations, 
free-streaming photons will keep high- and low- density regions
isothermal with baryonic pressure gradients remaining. 
In this regime, dissipation of inhomogeneities proceeds via
expansion of high density regions
towards low density regions and the concomitant transport of
material towards the annihilation region. The motion of the charged
particles, protons and light elements, is impeded by the Thomson
drag force, which acts on the electrons dragged along by the 
charged nuclei~\cite{Pe:71}. Balance between pressure forces and the
Thomson drag force yields a terminal velocity  
$v=\tot r_{100}/\tot t$~\cite{JF:94}, 
\begin{eqnarray}
v\approx \frac{3}{ 4\sigma_T\varepsilon_{\gamma}n_e}\frac{1}{R^2}
\frac{\tot P}{\tot r_{100}}\; ,
\end{eqnarray}
where $\tot P/\tot r_{100}$ is  radial (anti-) baryonic pressure gradient,
$r_{100}$   radial coordinate as measured on the comoving
scale, $\varepsilon_{\gamma}$  photon energy density, and
$n_e=n_{e^-}-n_{e^+}$  net electron density.
One finds for the pressure exerted by baryons and electrons
below $T\approx 30 $~keV
\begin{equation}
P \approx  \bar{T}\net\biggl(\sum_i\Delta_i+\Bigl(
{n^\ast_{\rm pair}}^2+\Bigl(\sum_iZ_i\Delta_i\Bigr)^2\Bigr)^
\frac{1}{ 2}\biggr)\; ,
\label{E:pressure}
\end{equation}
with the sum running over all nuclei $i$ with nuclear charge
$Z_i$. 
Note that expression~(\ref{E:pressure}) quickly reduces to the
pressure exerted by an ideal gas when the reduced $e^\pm$ pair density
$n^\ast_{\rm pair}=n_{\rm pair}/\net$ becomes negligible compared to
$\sum_iZ_i\Delta_i$~\cite{JF:94}. 
\section{Matter-Antimatter Annihilation}\label{S:nucl}
\subsection{Annihilation Reactions and Cross Sections}
The dominant process in nucleon-antinucleon interaction is direct
annihilation into pions,
\begin{equation}\begin{array}{lclllclclcl}
\multicolumn{3}{l}{
\left.\begin{array}{lcl}
p&+&\bar{p}\\
p&+&\bar{n}\\
n&+&\bar{n}\\ 
n&+&\bar{p}
\end{array}\right\}}&\multicolumn{4}{l}{\to \pi^0,\pi^+, \pi^- (\gamma,
\nu\bar{\nu})}.
\end{array}
\end{equation}
Electromagnetic annihilation ($p + \bar{p} \to \gamma + \gamma$) is
suppressed by a factor of $(m_e/m_p)^2 \approx 3 \times
10^{-7}$. Annihilation via the bound state of protonium is also
possible, but the cross section is smaller by $(m_e/m_p)^{3/2} \approx
10^{-5}$ compared to direct annihilation.
 
The charged pions  either decay with a lifetime of $\tau_{\pi^\pm}=
2.6 \times 10^{-8} \s$ directly into leptons\\ 
\setlength{\unitlength}{.5cm}
\begin{minipage}{\columnwidth}
\begin{eqnarray}
\pi^+ &\to&\mu^+ + \nu_\mu\\
&&\begin{picture}(1,1)(0,0)
\put(.2,0.2){\line(0,1){1}}
\put(.2,0.2){\vector(1,0){1}}
\end{picture} \hspace{.5em}e^+ + \nu_e + \bar{\nu}_\mu \nonumber\\
\pi^- &\to&\mu^-	+\bar{\nu}_\mu\\
&&\begin{picture}(1,1)(0,0)
\put(.2,0.2){\line(0,1){1}}
\put(.2,0.2){\vector(1,0){1}}
\end{picture} \hspace{.5em} e^- + \bar{\nu}_e + \nu_\mu\nonumber\;,
\end{eqnarray}
\end{minipage}

\begin{figure}[t]
\centerline{\psfig{figure=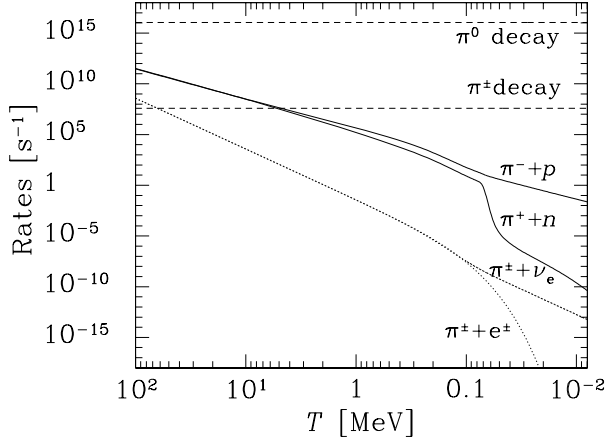,width=3.375in}}
\caption{Interaction rates for pions with leptons (dotted lines) and
        nucleons (full lines) as a function of cosmic
        temperature. Here a baryon-to-photon ratio 
        $\eta = 4\times 10^{-10}$ has
        been assumed. For
        comparison, the pion decay rates are also shown (dashed lines).
\label{F:pirates}}
\end{figure}

\bigskip
\noindent or may be transformed into $\pi^0$ via charge-exchange 
\begin{equation}
\begin{array}{lclclcl}
\pi^+ &+& n &\to& p &+& \pi^0\\
\pi^- &+& p &\to& n &+& \pi^0\\
\end{array}
\label{E:chex}
\end{equation}
or weak interactions 
\begin{equation}
\begin{array}{lclclcl}
\pi^\pm &+& e^\pm &\to& \nu_e/\bar{\nu}_e &+& \pi^0\\
\pi^\pm &+& \nu_e/\bar{\nu}_e &\to&e^\pm  &+& \pi^0\;.
\end{array}
\end{equation}
The neutral pions subsequently decay into photons with $\tau_{\pi^0}=8.4
\times 10^{-17} \s$.
The rates for the three channels, decay, charge exchange, and weak
interactions are shown in Fig.~\ref{F:pirates}. Below a temperature of
a few MeV,  decay dominates the loss of charged pions unless the local
baryon-to-photon ratio is well in excess of $\eta \sim 10^{-10}$. 
Nevertheless, even at low $\eta$ some
pions may charge-exchange on nucleons; possible consequences thereof
will be discussed below.  Weak interactions with the ambient leptons
do not significantly contribute to the pion interaction rates in the
temperature range relevant for our work ($T \lesssim 20
$~MeV).  Neutral
pions never have a chance to interact with either leptons or nucleons,
due to their rapid decay.

Annihilation of antinucleons on light nuclei, $N$, produces a wealth of
secondary particles,
\begin{eqnarray}
\bar{p} /\bar{n} + N &\to& N^{'},p,n,\pi\;.
\end{eqnarray} 
In Table~\ref{T:probab} we give probabilities for the production of
secondary nuclei $N^{'}$ in the $^4$He annihilation process
following Ref.~\cite{BBB:88}. Production of secondary nuclei during
antinucleon annihilations on the other light nuclei, and
anti-$^4$He annihilation on $^4$He, are relatively unimportant for this
study. Due to the huge difference in the abundances of \he4 on the
one hand and the other nuclei on the other hand, the destruction of
only a minute fraction of \he4 may already have significant impact on
the abundances of the other primordial elements. Disruption of the
other elements will occur with much smaller probabilities
due to the smaller abundances of these isotopes, and the secondaries
of these processes will never contribute significantly to the
respective abundances. In our numerical calculations we assumed that
disruption of all elements but \he4 results in free nuclei only.

\begin{table}
\caption{Probabilities for the production of secondary nuclei and nucleons
in \pbar\he4 annihilations, derived from the branching ratios given in
Ref. \protect\cite{BBB:88}.
\label{T:probab}}
\begin{tabular}{lllll}
$P_n$&$P_p$&$P_{^2{\rm H}}$&$P_{^3{\rm H}}$&$P_{\rm ^3He}$\\
\hline
0.51&0.28&0.13&0.43&0.21\\
\end{tabular}
\end{table}

We are interested in annihilations of antimatter domains between 
shortly before the epoch of weak freeze out and recombination such
that the
annihilation cross sections for thermal nucleons with kinetic energies
between a few MeV and about $10^{-7}$~MeV are needed.
Experimental
data are available only down to an incident momentum of about 30--40
MeV, corresponding to kinetic energies of about 1~MeV. Therefore, we
have to extrapolate the experimental data with the help of existing 
theoretical calculations for the cross sections down to the
relevant energy range.  At such low energies, the Coulomb forces
between charged particles become important, thus systems with Coulomb
interactions like \pbar$p$\ and \pbar$N$, and without, like \nbar$n$,
\nbar$p$, and \nbar$N$, have to be treated separately.

The product of annihilation cross section $\sigma_{\rm ann}$ and
relative velocity $v$ in systems with at least one neutral particle is
known to be approximately constant at low energies
\cite{CPZ:97}. Experimental values of $\sigma_{\rm ann} v = 40\pm3 \,
{\rm mb}\,c$ and $32\pm5\, {\rm mb}\,c$ were obtained at center of
mass momenta of 22 MeV/$c$ and 43 MeV/$c$, respectively~\cite{MCK:88}.
In our calculations, we used a constant value of $\sigma_{\rm ann} v =
40 \mb\, c$.
  
In systems with Coulomb interactions, such as the $\bar{p}p$ or the
$\bar{p}N$ system, the behavior of the low-energy annihilation
cross section is drastically modified due to Coulomb attraction. 
Indeed, the charged particle
low energy annihilation cross section is found to be inversely
proportional to the square of the incident momentum and therefore the
reaction rate is formally divergent at zero energy. (This divergence
is of course removed when matter and antimatter reach atomic form at
low temperatures.)  
Again, there are no
experimental data below about 1~MeV kinetic energy.  The available
experimental data at higher energies is however well reproduced by
phenomenological calculations~\cite{CP:93,CP:96}. 
The results of these calculations depend on the mass and the charge of the
nucleon/nucleus under consideration as well as on some phenomenological
parameters, which may be determined experimentally. 

For energies below about $10^{-1} $~MeV, which are mostly relevant for
our study, the thermally averaged $\bar{p}p$ annihilation rate 
in a plasma of temperature $T$ may be approximated by
\begin{eqnarray}
<\sigma_{\rm ann} v>\, \approx f_{\bar{p} p} \mb\, c \
\sqrt{\frac{\MeV}{T}}\; , 
\label{E:cross_p_pbar}
\end{eqnarray}
where $f_{\bar{p}p}=32$ is a numerical constant.
The equivalently defined numerical constants  for the $\bar{p}$N
annihilation channels are 
of similar magnitude, $f_{\bar{p}^2{\rm H}}\approx f_{\bar{p} ^3{\rm
H}}\approx 16$ and $ f_{\bar{p}^3{\rm He}}\approx f_{\bar{p}^4{\rm
He}}\approx 20$. 
These were obtained  using the  results of the
phenomenological calculations as given in Refs.~\cite{CPZ:97,PBRZ:00}
based on the experimental data as given in Refs.~\cite{Zetal:00ab}.

\subsection{Impact of Secondaries in Nucleon-Antinucleon Annihilations on BBN}
\label{S:second_nucleon}
It is of interest if annihilation-generated photons and pions, or
their decay products, may alter the abundance yields, either through
their effect on weak freeze out or by, for example,
photodisintegration or charge exchange reactions. In a single
annihilation event, about 5--6 pions with momenta ranging from tens to
hundreds of MeV are produced. Reno \& Seckel~\cite{RS:88} showed that
prior to $e^\pm$ annihilation the thermalisation time scale for
charged pions is always  shorter than the hadronic interaction time. 
At later times, the charged pions have no chance to interact due to their
short lifetime.
We may thus use the thermally averaged  charge exchange cross
sections (cf. Eq.~\ref{E:chex}) given in Ref.~\cite{RS:88} to calculate 
the ratio between a typical
charge exchange interaction time and $\tau_{\pi^\pm}$
\begin{eqnarray}
\frac{\tau_{\rm cex}}{\tau_{\pi^\pm}}
\approx 0.01\,
\left(\frac{\eta_{\rm local}}{4 \times
10^{-10}}\right)^{-1}\,\left(\frac{T}{\MeV}\right)^{-3}.  
\label{E:cex}
\end{eqnarray} 
Charged pions may thus only charge exchange, if $\eta_{\rm local} \gg
10^{-10}$, and the temperature is not much lower than 1~MeV. Due to
their electromagnetic interaction, charged pions remain mainly
confined to the annihilation region as long as $e^\pm$ pairs are still
abundant. Within that region, $\eta_{\rm local}$ is typically much
lower than the \mbox{(anti-)} baryon-to-photon ratios anywhere else due to
prior matter-antimatter annihilation (cf.~App.~\ref{S:ann_regio}).
At lower temperatures, when the pions may easily move within the
primordial fluid, charge exchange reactions are negligible due to the
small nuclear densities, as is apparent from Eq.~(\ref{E:cex}). We
therefore observe only negligible impact of charge exchange reactions
on the BBN abundance yields in our numerical simulations, even for
early matter-antimatter annihilation and large $R_A$ (implying large
$\eta_{\rm local}$ in matter- and antimatter- domains).

The leptonic secondaries $\mu^\pm, e^\pm$ and $\nu$ do not
modify the details of weak freeze out, unless the  number of
annihilations per photon is extremely large. As long as this number is
not approaching unity, annihilation generated $\nu_e$'s have negligible effect 
on the $n/p$-ratio, since their
number density is  orders of magnitude smaller than that of the
thermal $\nu_e$'s, which govern the weak equilibrium. The same holds for 
electrons and positrons produced in $\mu$-decay which are quickly thermalized
by  electromagnetic interactions. 

In each annihilation event about half of the total energy is released 
in form of electromagnetic energy.
Important impact on the BBN abundances may result through
these annihilation-generated $\gamma$-rays and $e^{\pm}$
cascading on the background photons
(and on pairs before $e^\pm$ annihilation) via pair production and
inverse Compton scattering on a time scale rapid compared to the time
scale for photodisintegration of nuclei~\cite{Li:80,EGLNS:92,PSB:95}.
After cosmic $e^\pm$ annihilation, the cascade only terminates when
individual photons do not have enough energy to further pair-produce
on background photons.  For temperatures $T \,\gtrsim\, 5 $~keV, the
energy of $\gamma$-rays below the threshold for $e^{\pm}$-production
does not suffice for the photodisintegration of nuclei. If
annihilations occur below 5~keV, the light nuclei gradually become
subject to photodisintegration, according to their binding energy.

Destruction of \h2, \h3 and
\he3 by photodisintegration is thus possible for temperatures below
$T \lesssim T_{\gamma ^2{\rm H}}\approx 5 $~keV and $T \lesssim T_{\gamma {\rm
^3He}}\approx T_{\gamma ^3{\rm H}} \approx 2 
$~keV, respectively. This destruction is nevertheless subdominant to
the production of these isotopes by 
photodisintegration of \he4, possible at lower temperatures 
($T_{\gamma {\rm ^4He}} \simle 0.4$~keV), simply due to the larger
abundance of $^4$He. Thus photodisintegration of
\h2, \h3 and \he3 may only be important if annihilations
take place in the temperature range $0.4\,{\rm keV}\simle T \simle
5$~keV. However, direct production of \h2, \h3 and \he3 via
annihilations on \he4 dominates destruction by
photodisintegration (see Sec. 
\ref{S:sec:nuclei} below). The photodestruction
factor for, e.g. \h2, may be roughly estimated from the 
photodestruction factor for \he4 given in Ref.~\cite{PSB:95}.
Since the cross section for
the competing process, i.e. Bethe-Heitler pair production on protons,
does only slowly vary with temperature, these destruction factors 
may be used in
the relevant temperature range, $5 \keV \lesssim T_{\rm ann} \lesssim
0.4 $~keV, but have to be scaled by the target abundances. Thus one
estimates a fraction of about $0.1\, (\mhe4/^2{\rm H}) \approx 10^{-4}$
\h2 nuclei per GeV electromagnetic energy injected to be
photodisintegrated. Direct annihilations will create about
$0.2\, (\mhe4/p) \approx 10^{-2}$ \h2 nuclei per
annihilation. Even though a fraction of those will thermalize within
the annihilation region, and thus be subsequently annihilated, 
it is well justified to neglect photodisintegration of lighter
isotopes. Photodisintegration of nuclei lighter than \he4 was thus
not taken into account in our simulations.

For the \he4 photodisintegration yields we used the results
given in Protheroe {\it et al.}\cite{PSB:95}.
Production of \he3 and $^3$H exceeds the \h2 yield by a factor of ten,
thus typically producing a large $^3{\rm He}/^2{\rm H}$ ratio.
The photodisintegration yield for \he3 is peaked at a
temperature of $T \approx 70$~eV and becomes
significantly smaller at lower temperatures.
When photodisintegration of
\he4 occurs it creates initially energetic \he3 and \h3 nuclei.
The importance of the energetic $^3$H and \he3 nuclei resulting 
from photodisintegration is twofold; they do
not only directly increase the \he3 abundance, but may also lead to
production of \li6 via $^3$H/\h3 + \he4 $\to$ \li6 + $n/p$, as was recently 
stressed by one of us~\cite{Je:00}. We take the \li6 yields as given
in Ref.~\cite{Je:00} (for more details, cf. 
Sec.~\ref{S:sec:nuclei} below).  

Note that all photodisintegration yields have been calculated using
the generic $\gamma$-ray spectrum given in Protheroe {\it et al.}
\cite{PSB:95}.  This procedure is only adequate as long as the energy of the
injected photons, $E_\gamma \approx 200 $~MeV, is beyond the threshold
for $e^{\pm}$-pair production of the injected $\gamma$'s on the
background photons, $E_C(z) \approx 4.7 \times
10^7 (1+z)^{-1} $~MeV, i.e. for $z \gtrsim 2 \times 10^5$,
corresponding to $T_{\rm ann} \gtrsim 50\,$eV. The results for 
later annihilation, i.e for antimatter domains
on scales larger than $\ra \gtrsim 10^3$~cm thus have to be
interpreted with some care. In order to be able to give conservative
limits, we have done simulations where photodisintegration was ignored
which gave weaker limits by a factor of a few (see
Fig.~\ref{F:limits}).  

\subsection{Impact of Secondaries in Antinucleon-Nucleus Annihilations on BBN}
\label{S:sec:nuclei}
While we expect secondaries of nucleon-antinucleon
annihilations to have only significant effect at lower temperatures, 
energetic nuclei arising in  antinucleon-nucleus annihilations 
may substantially modify the
light element abundances for temperatures as high as
$T < T_{{}^4{\rm He}}\approx 80\,$keV. This may occur
through direct production of light isotopes
in antinucleon annihilations on $^4$He as well as 
through possible subsequent non-thermal
fusion of these energetic light isotopes on $^4$He.
Since the \he4 abundance exceeds the
abundances of the other isotopes by orders of magnitude, $\bar{p}$\he4
is the dominant antinucleon-nucleus annihilation process. The
relative probabilities for the production of the various 
secondary nuclei arising in  \he4 disruption are given in
Table~\ref{T:probab}. The secondary nuclei are produced within the
annihilation region and may thus themselves be subject to
annihilation, unless they are able to escape from the annihilation
region. On average, the secondary nuclei gain a kinetic energy $E_0$  of a
few tens of MeV~\cite{BBB:88}. Their transport is then initially
described by free-streaming until their energy has decreased 
to thermal energies through interactions with the plasma, after which
transport has to proceed via thermal diffusion.  
The dominant energy loss mechanisms for energetic charged nuclei 
in a plasma with kinetic energy below 1~GeV 
are plasmon excitations and Coulomb scatterings. Note that these 
processes have negligible impact
on the direction of the momentum of the energetic nuclei such
that free-streaming is a good approximation \cite{Ja:75}.
The distance covered until the
kinetic energy of the particles has decreased to the thermal energy of the
plasma defines the stopping length,
\begin{eqnarray}
l^{\rm stop}=\int_0^{l^{\rm stop}}\tot x =
\int_{E_0}^{E_{\rm thermal}} \frac{\tot x}{\tot E} \tot E\; .
\label{E:l_stop}
\end{eqnarray} 
Provided $l^{\rm stop}$ is larger than the size of the annihilation
region, all nuclei which become thermalized in a matter domain
(typically about 1/2) have a good chance to survive.  
If we calculate the energy loss per distance following Ref.~\cite{DEHS:88}
we find for the stopping length measured in our comoving coordinates
\begin{eqnarray} 
l^{\rm stop}_{100}&\approx&55  \, \cm \,  \left(\frac{{\rm
keV}}{T} \right)^2  
\left(\frac{E_0}{50\,{\rm MeV}}  \right)^2 \nonumber\\
&&\times\left(\frac{4 \times 10^{-10}}{\eta_{\rm local}}  \right)
Z^{-2}\; ,
\label{E:lstop_ch2}
\end{eqnarray}
where $Z$ is the charge of the energetic nucleus.
In an analogous manner we may calculate a stopping time 
\begin{eqnarray}
\tau_{\rm stop}=\int_0^{E_{\rm thermal}}
\frac{1}{v(x)}\frac{\tot x}{\tot E}\tot E
\label{E:tau_stop}
\end{eqnarray} 
needed to slow down a
particle to thermal energies. Evaluation of
the integral yields for charged particles
\begin{eqnarray}
\tau_{\rm stop}^p
&\approx& 
3.25\s \, \left(\frac{{\rm keV}}{T} \right)^3 
\left(\frac{E_0^{3/2}-E_{\rm thermal}^{3/2}}{(50\,{\rm MeV})^{3/2}}
\right)  \nonumber\\
&&\times\left(\frac{4 \times 10^{-10}}{\eta_{\rm local}}  \right)
Z^{-2}\; .
\label{E:tau_stop_ch}
\end{eqnarray}	

Neutrons lose their energy through nuclear scatterings. In contrast to
the charg\-ed particle interactions discussed above, the deflection
angle in a nuclear scattering event may be large, such that the use
of Eq.~\bref{E:l_stop} is inappropriate since it relies on the
free-streaming assumption. Rather,
the distance covered by the neutrons is described by a random
walk. The stopping time is nevertheless described by
Eq.~\bref{E:tau_stop}, since the energy loss does not depend on the
direction of the motion.  The energy loss per unit distance for neutrons
may be estimated via
\begin{eqnarray}
\frac{\tot E}{\tot x}= (-\ln f) E \sigma_{np} n_p\,,
\label{E:e_loss_n}
\end{eqnarray}
where $f$ is an approximate average
fractional energy loss in each elastic neutron-proton
scattering event. If we
assume a simple power law for the neutron-proton cross section $\sigma_{np}
\approx 10^3 \, {\rm mb} \, (E/10 \MeV)^{1.15}$
(cf. Fig.~1 in Ref.~\cite{Me:72}) and an
energy loss of 80 \% in each scattering event, we find
\begin{eqnarray}
\tau_{\rm stop}^n&\approx&1.56 \times 10^2 \s \,\left(\frac{
\keV}{T}\right)^3 \left(\frac{4 \times 10^{-10}}{\eta_{\rm
local}}\right)\nonumber\\ 
&&\times\left(\frac{E_0^{\,0.65}-E_{\rm thr}^{\,0.65}}{(10 \MeV)^{\,0.65}}\right).
\label{E:tau_stop_n}
\end{eqnarray} 

We may now compare $\tau_{\rm stop}$ for neutrons and protons with a 
typical \he4 spallation time scale, $\tau_{\rm sp} = \left(\langle 
\sigma_{\rm sp} v \rangle n_{^4{\rm He}} \right)^{-1}$. We find that only 
about a fraction of $10^{-3}$ 
of all energetic protons may spallate 
additional \he4. Since direct production of energetic protons and light nuclei 
in $\bar{p}$\he4 disruption is of similar magnitude, this effect may 
safely be neglected. For the energetic neutrons, we find that about 30 \% 
may spallate \he4. Since we obtain energetic neutrons in about half of 
the annihilation events, additional \h3 or \he3 will be produced 
in about a tenth of the $\bar{p}$\he4 annihilations. This is not significant 
compared to the secondary \h3 and \he3 produced directly in the 
$\bar{p}$\he4 annihilations with a probability of about 60\% and will 
therefore be neglected in the numerical treatment.

The annihilation generated energetic
light nuclei may however be important as a source for very rare
light elements such as $^6$Li via the fusion reactions  \h3 + \he4 
$\to$ \li6 + $n$ and \he3 + \he4 $\to$ \li6 + $p$~\cite{remarkLi}.
Using a value of 35~mb for the fusion cross section, the threshold energies of 
$E_{\rm th} = 4.80$ MeV and $E_{\rm th} = 4.03$ MeV, respectively, and the 
energy distribution for the nonthermal mass three nuclei as given in 
\cite{BBB:88}, we find 
\begin{eqnarray}
\langle P_{^3{\rm H}^4{\rm He} \to n^6{\rm Li}}\rangle \approx 2 \times 10^{-6}
\end{eqnarray}
and
\begin{eqnarray}
\langle P_{{^3{\rm He}}^4{\rm He} \to p^6{\rm Li}}\rangle \approx 5
\times 10^{-7} 
\end{eqnarray}
for the probabilities to produce \li6 from energetic mass three nuclei.
The calculation is done similarly to the ones in
Refs.~\cite{DEHS:88,Je:00}, where energy loss of 
energetic nuclei according to Eq.~(\ref{E:tau_stop_ch}) has been taken
into account.
The number of \li6 nuclei produced per antiproton annihilation is thus
\begin{eqnarray}
N_{^6{\rm Li}} &\approx&  (P_{^3{\rm H}}
P_{^3{\rm H}^4{\rm He} \to n^6{\rm Li}}+  P_{^3{\rm He}}P_{{^3{\rm
He}}^4{\rm He} \to p^6{\rm Li}}) \\ 
&&\times\;\left(\frac{\sigma_{\mpbar^4{\rm He}}}{\sigma_{\mpbar p}}\right)
\left(\frac{n_{^4{\rm He}}}{n_p}\right)\nonumber\\
&\approx& 1.8 \times 10^{-8}
\left(\frac{Y_{\rm p}}{0.25}\right), \nonumber
\end{eqnarray}
where $P_{^3{\rm H}}$ and $P_{^3{\rm He}}$ are the probabilities to create
\h3
or \he3 in a \pbar\he4 annihilation event (see Tab.~\ref{T:probab}). 
A simple estimate for the total synthesized 
\li6/H abundance (excluding production via \he4 photodisintegration)
is thus   
\begin{eqnarray}
{^6{\rm Li}\over {\rm \, H}}\approx  \biggl({n_{\bar{b}}\over
n_{b}}\biggr)N_{^6{\rm
Li}}\approx 1.8 \times 10^{-9} 
\left(\frac{\RA}{0.1}\right)
\left(\frac{Y_p}{0.25}\right)\, ,\label{6li}
\end{eqnarray}  
for $\RA\ll 1$ and where it is understood that only that fraction of
antimatter
has to be inserted in Eq.~(\ref{6li}) 
which has not annihilated by temperature 
$T_{^4{\rm He}}\approx 80\,$keV.
This is many orders of magnitude higher than the standard BBN value, 
$n_{^6{\rm Li}}/n_p = {\cal O}(10^{-13})$ and will therefore provide very
stringent limits in some areas of the parameter space, as will be
discussed in Sec.~\ref{S:Res}.

\begin{figure}
   \centerline{\psfig{figure=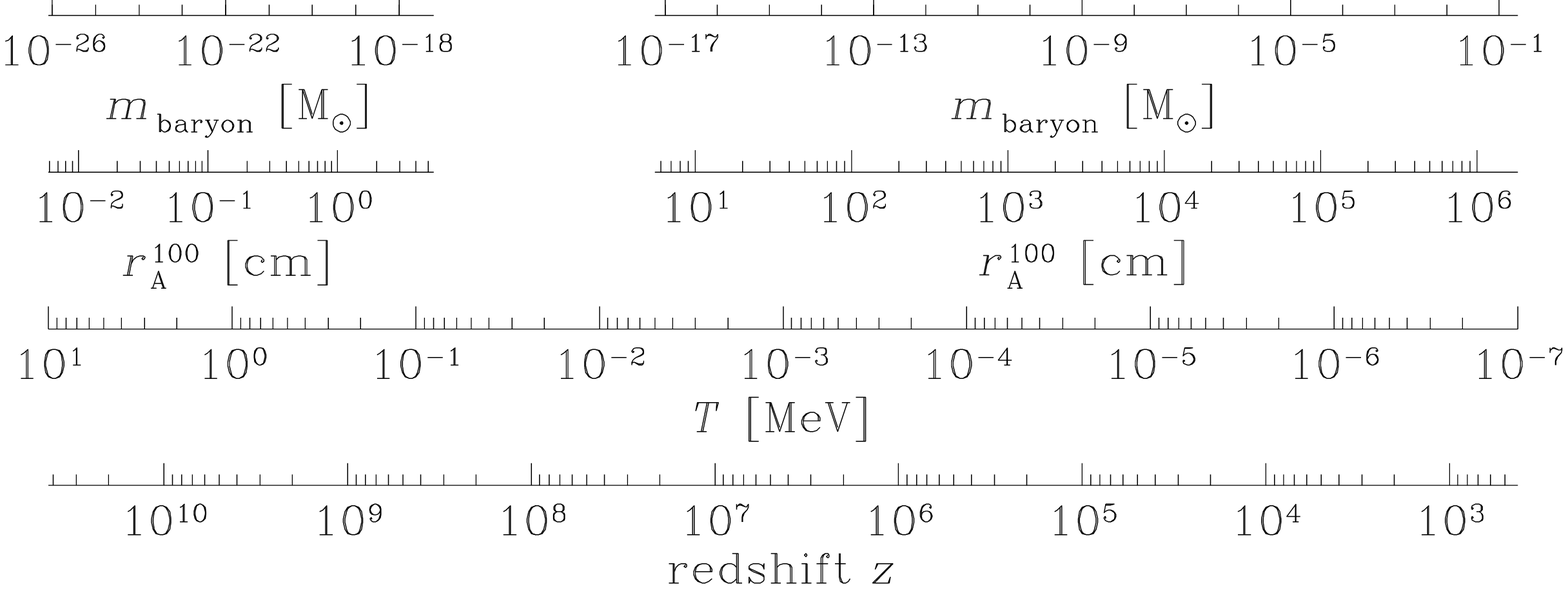,width=3.375in}}
{\caption{Spatial dimension of the antimatter regions and corresponding
        annihilation temperature for an antimatter fraction of $\RA =
        0.1$. Also given is the baryonic mass in 
        units of the solar mass $M_\odot$ 
        contained in the antimatter region and the annihilation
       redshift. Note that annihilation is stalled in the temperature
         range $T_{^4{\rm He}} \gtrsim T \gtrsim 5 \keV$, as is
	explained in the text. 
\label{F:scales}}}
\end{figure}

\section{BBN with Matter-Antimatter Domains}\label{S:bbnma}
After having discussed the different dissipation mechanisms
of antimatter domains in the early Universe 
as well as the annihilation reactions and the possible impact of 
annihilation generated secondaries
on BBN, we are now in a
position to put all this together in order to examine 
the influence of annihilating antimatter domains in the early universe
on the BBN light element abundance yields. Clearly, such scenarios involve
such a multitude of nuclear reactions and hydrodynamic dissipation
processes that obtaining fairly accurate predictions for the BBN
yields requires numerical simulation. We have
therefore substantially modified the inhomogeneous BBN code by
Jedamzik, Fuller \& Mathews~\cite{JFM:94}, originally including
nuclear reactions, baryon diffusion, and fluctuation dissipation by
photon and neutrino induced processes, as to also include nuclear reactions
between antimatter, matter-antimatter annihilation reactions, 
free-streaming of secondary nuclei produced in annihilations, the
non-thermal
fusion reactions of secondaries, as well as photodisintegration of
$^4$He through annihilation generated cascade $\gamma$-rays. Some
processes
are not included in our simulations, due to their marginal impact on
BBN as outlined in the last section. For more
details on the procedure of the numerical simulation 
the reader is also referred to 
App.~\ref{S:num}.

A detailed analytic and numerical analysis of the actual structure of
the annihilation region, i.e. the region at the domain boundaries
where the bulk of annihilations occur, is presented in
App.~\ref{S:ann_regio}. Our simulations do not accomplish to
resolve the physical width, $l_{\rm ann}$, of this region due to the
increasingly large ratio, $R_A/l_{\rm ann}$, and the required extreme
dynamic range to resolve both scales. However, we believe, as we argue
in App.~\ref{S:ann_regio}, that this \lq\lq flaw\rq\rq\ of our simulations has
only little impact on the accuracy of our results.  This is also
confirmed by the relative independence of our results on the total
number of zones employed in the simulations (see App.~\ref{S:num}).
\begin{figure*}[t]
   \centerline{\psfig{figure=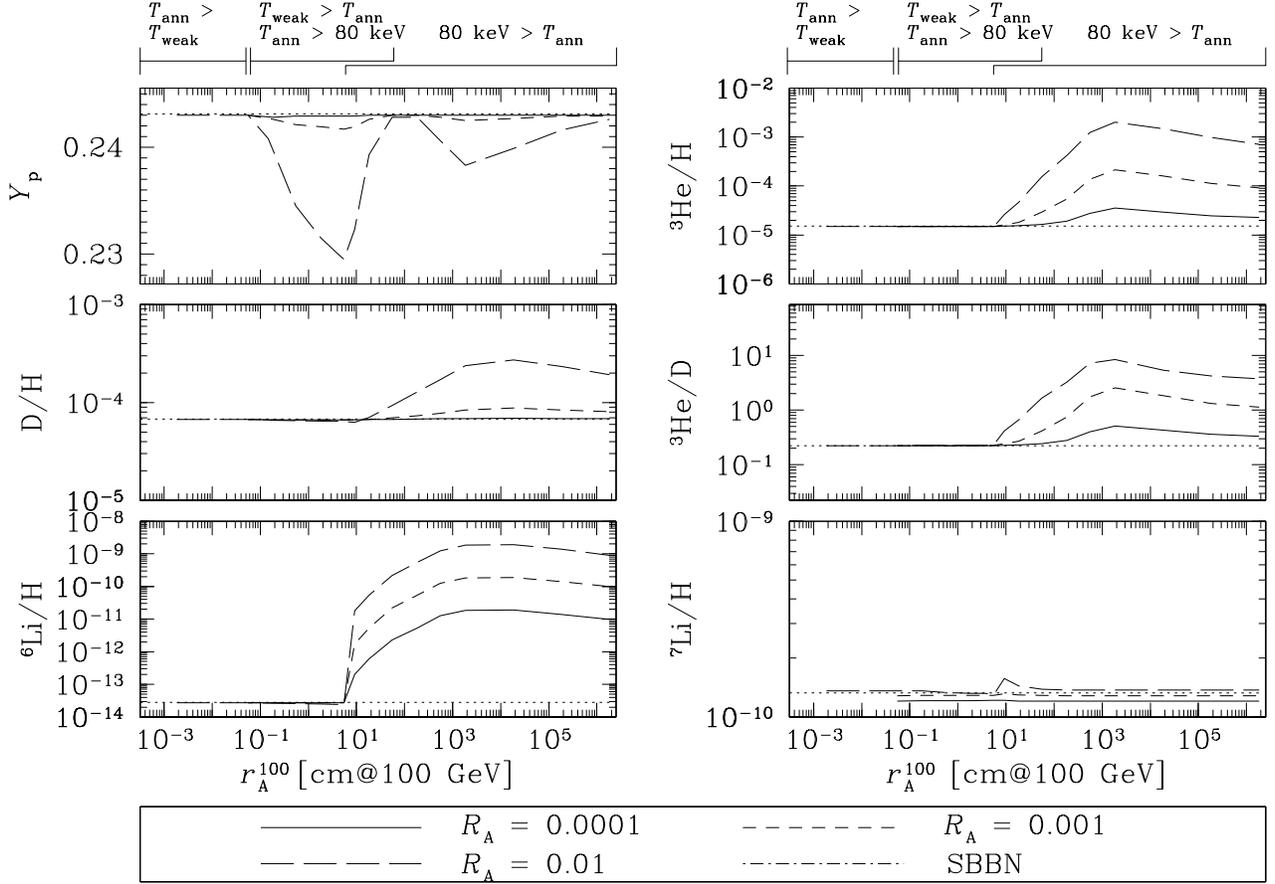,width=\textwidth}}
{\caption{Results of BBN calculations in a Universe with antimatter
        domains. Shown are the \he4 mass fraction, the abundances
        of \h2, \he3, \li6 and \li7 relative to hydrogen and the ratio
        of \he3 over \h2 as a function of antimatter domain radius, 
        $r_A^{100}$, for several low values of the
        antimatter-to-matter
        ratio, $\RA$  (see legend).  
 \label{F:comb_1}}}
\end{figure*}

\begin{figure*}[t]
   \centerline{\psfig{figure=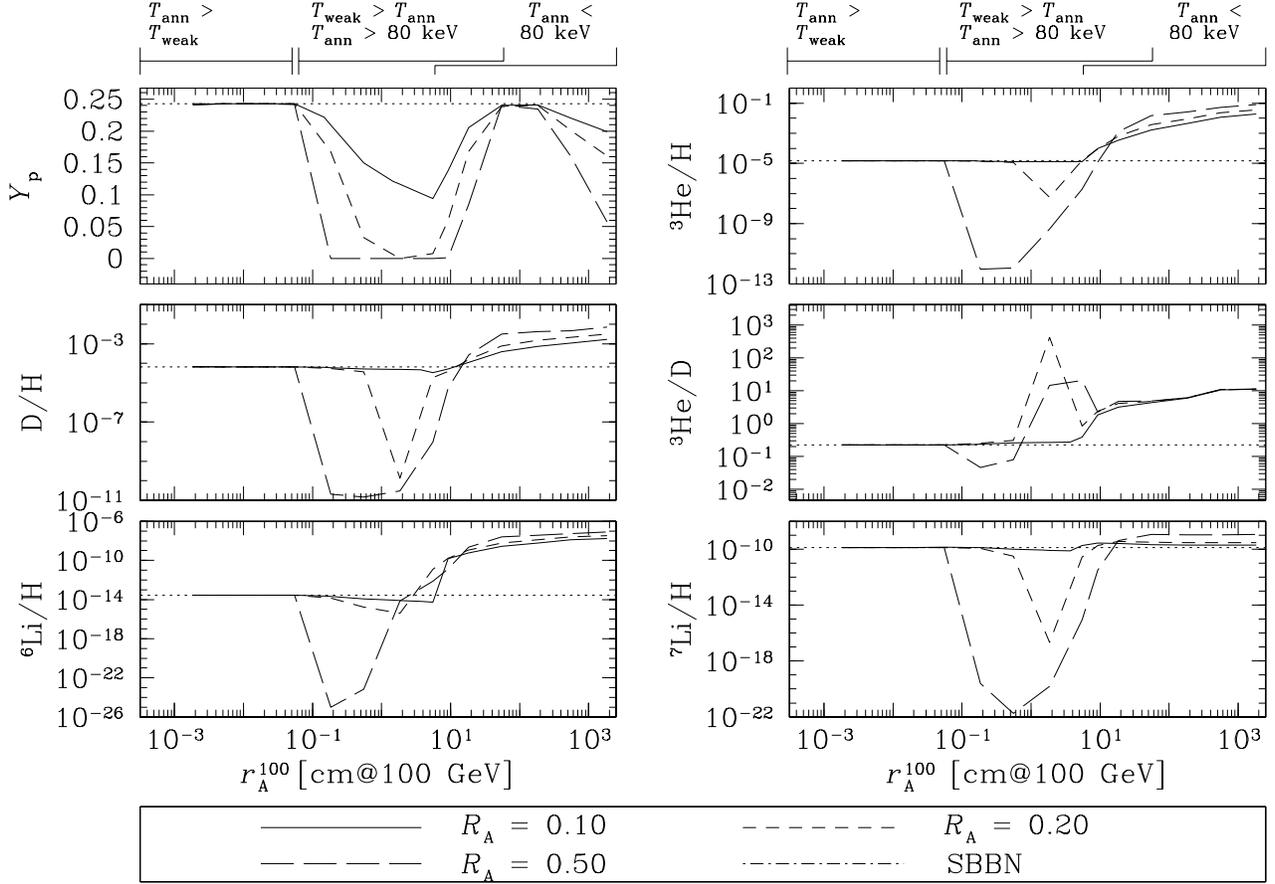,width=\textwidth}}
{\caption{As Fig.~\ref{F:comb_1} but for higher antimatter-to-matter
ratios, $R_A$.
 \label{F:comb_2}}}
\end{figure*}
The relationship between antimatter domain size, $r_A^{100}$, and
approximate annihilation temperature $T$ and redshift $z$ of a domain
is shown in Fig.~\ref{F:scales} and is determined by neutron diffusion
at early times and hydrodynamic expansion at late times. Since
dissipation by neutron diffusion for temperatures $T\, \simge\, T_{\rm
^4He}\approx 80\,$keV is relatively more efficient than dissipation by
hydrodynamic expansion at somewhat lower temperatures, antimatter
domain annihilation does typically not occur in the temperature regime
between $T_{\rm ^4He}$ and  5~keV. This implies that
\lq\lq injection\rq\rq\ of antimatter and annihilation between the
middle and the end of the BBN freeze-out process, as envisioned in the
scenarios of Ref.~\cite{DY:87_YD:88}, may not operate in scenarios
with a matter-antimatter domain structure in the early universe.

We present the detailed numerical results of our study in
Figs.~\ref{F:comb_1} and~\ref{F:comb_2}. Shown are the abundances of
the respective elements and the \he3/\h2 ratio for a number of
matter-antimatter ratios $\RA$ as a function of the typical size of
the antimatter regions $\ra$.  In the subsequent discussion of our
results we will distinguish between three different limiting cases,
according to the segregation scale of matter and antimatter, or
equivalently, the approximate matter-antimatter annihilation time.
The fractional contribution of baryons to the critical  density of the
Universe, $\Omega_b$, was kept at a constant value of
$\Omega_b= 0.0125 h_{100}^{-2}$ in all simulations, where
$h_{100}$ parameterizes 
the value of the Hubble parameter $H_0$ today, $h_{100}=
H_0/(100\, \rm{km\, s^{-1} \,Mpc^{-1}})$.  
The corresponding  value of the SBBN parameter is $\eta= 3.4 \times
10^{-10}$.

\subsection{Annihilation Before Weak Freeze Out}
\begin{figure}
 \centerline{\psfig{figure=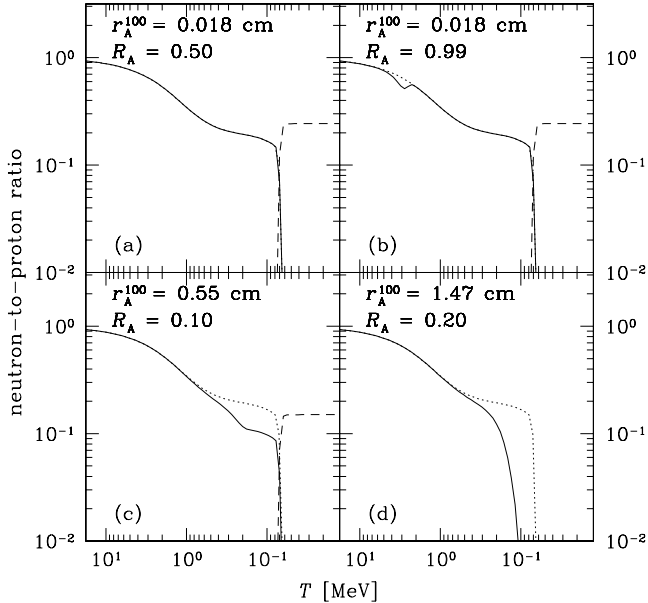,width=3.375in}}
{\caption{The neutron-to-proton ratio (solid line) and the \he4 mass
        fraction (dashed line) as a function of
        temperature for 
        different sizes of the antimatter domains, $\ra$, and different
        values of the matter-to-antimatter ratio, $\RA$, 
        as indicated in each panel. For comparison, the
        dotted line shows the unperturbed neutron-to-proton
        ratio in a Universe without antimatter.
 \label{F:np}}}
\end{figure}

The key parameter which determines the primordially synthesised amount
of \he4 is the $n/p$-ratio at a temperature of $\The
\approx 80$~keV. Early annihilation of antimatter proceeds mainly via
neutrons diffusing towards the antimatter domains and being annihilated
at the domain boundaries whereas antineutrons diffusing towards the
matter domains may annihilate on both, neutrons and protons. 
The net effect of both processes is the preferential annihilation of 
antinucleons on neutrons, perturbing the $n/p$-ratio towards smaller
values as detailed in Ref.~\cite{RJ:98}. (One finds this effect often
to be even more pronounced since after some initial
annihilation of protons close to the domain boundary the annihilation
region is completely void of protons.) If these perturbations in the 
$n/p$-ratio persist down to $T_{\rm ^4He}$, a significant reduction in the 
synthesized $^4$He mass fraction may result. A potential effect in the
reverse direction, increase of the $n/p$-ratio by preferential
pion-nucleon charge exchange reactions on protons (cf. Eq.~\ref{E:chex}), 
is subdominant as discussed in Sec.~\ref{S:second_nucleon}.
It is of interest at which temperatures these possibly large
perturbations in the $n/p$-ratio may still be reset by proton-neutron
interconversion via weak interactions governed by the rate
$\Gamma_{\rm weak} \approx G^2_{\rm Fermi} T^5$.
In the upper two panels of
Fig.~\ref{F:np} we show the $n/p$-ratio as a function of
temperature for comparatively early antimatter domain annihilation. 
In panel (a), the antimatter fraction was chosen to be
$\RA = 0.5$ and the length scale of the antimatter regions to be $\ra
= 0.018$~cm, corresponding to an approximate annihilation temperature
of about 5 MeV. The $n/p$-ratio for this parameter combination is observed
to be virtually indistinguishable from the $n/p$-ratio in a standard
BBN (SBBN) scenario.
Thus the final \he4 mass fraction (dashed line) coincides with the
SBBN value of $Y_p \approx 0.24$. Only when the matter-to-antimatter
ratio rises to values of order unity (panel b), the $n/p$-ratio in the
presence of antimatter annihilations
deviates significantly from the corresponding quantity in a SBBN
scenario
(shown by the dotted line). However,
after the annihilation has completed, the weak interactions are still rapid
enough to reestablish weak equilibrium and thus the final \he4 mass
fraction and the other light element abundances emerge unaffected.
In Figs.~\ref{F:comb_1} and~\ref{F:comb_2} one observes that significant impact on the
synthesized $^4$He mass fraction occurs only for $\ra$ larger than
$5\times 10^{-2}$cm to $10^{-1}$cm, depending on $R_A$.

\subsection{Annihilation After Weak Freeze Out, But Before \he4 Synthesis}
The situation changes when annihilation occurs during or after weak freeze out,
when neutron-proton interconversion ceases to be efficient.
Annihilation continues to proceed mainly via neutrons and
antineutrons, since proton diffusion is still hindered by the abundant
$e^\pm$ pairs.  However, neutrons and antineutrons which have diffused
out of their respective regions may now not be reproduced
anymore. Antimatter regions of typical size larger than the neutron
diffusion length at the cosmological epoch of weak freeze out
thus provide a very efficient sink for neutrons.  The $n/p$-ratio is
strongly affected in such models as is apparent from the lower two
panels of Fig.~\ref{F:np}.  In a model with, e.g., $\RA=0.1$ and $\ra
= 0.55\,$cm, annihilation proceeds at a temperature of $\approx 0.8$
MeV (panel c). At this temperature, the weak interactions are not
rapid enough to reestablish the equilibrium $n/p$-ratio. Thus the
final \he4 mass fraction $Y_p$ will be decreased compared to its SBBN
value. This is also evident from Figs.~\ref{F:comb_1}
and~\ref{F:comb_2} for antimatter domains with length scales between
$\sim 5\times 10^{-2}\,$cm and about $6\,$cm. For small antimatter
fractions, $R_A \lesssim 0.1$, the other light element abundances are
comparatively less affected than $^4$He, only for larger values of
$R_A$ production of \h2, $^3$He, and $^7$Li is also strongly suppressed.

Since the primordial \he4 mass fraction mostly depends on the $n/p$-ratio
at $\The$, it may be estimated analytically whenever $n/p|_\The$
is known,
\begin{eqnarray}
Y_p\approx \left.\frac{2(n/p)}{1+(n/p)}\right|_{\The}. \nonumber
\end{eqnarray}
If one assumes that annihilation occurs instantaneously, the
$n/p$-ratio in a scenario with annihilating antimatter domains may be
estimated by
\begin{eqnarray}
\left.\frac{n}{p}\right|_{\The}
 &\approx&  \frac{(n_0 /n_b) \, \exp\left[-{\Delta
t_1/\tau_n}\right]  - x \RA} 
{(p_0/n_b) - (1-x) \RA} \nonumber\\
&&\times\exp\left[-\Delta t_2/\tau_n\right]\; ,\label{E:anni_est}
\end{eqnarray} 
where $x$ is the fraction of antibaryons annihilating on neutrons,
$n_0$ and $p_0$ are the (pre-annihilation) neutron and proton
densities at $T\approx 0.2 
$~MeV, and $n_b$ is the actual baryonic density in the matter region.
Neutron decay is taken into account 
by the two exponentials, where $\Delta t_1$
is the time interval between the moment after which the neutron fraction is
(apart from annihilations) only affected by neutron decay ($T\approx
0.2~$MeV) and the moment of annihilation, while $\Delta t_2$ is the time
remaining until neutrons are incorporated into
\he4 at $\The \approx 80~$keV. 
Thus the two limiting cases between which this estimate should hold
are identified by $\Delta t_1\approx0 \, {\rm s}$, $\Delta t_2\approx 130
\, {\rm s}$ (annihilation at $T\approx 0.2 $~MeV) and $\Delta
t_1\approx 130 \, {\rm s}$, $\Delta t_2\approx 0 \, {\rm s}$
(annihilation at $T\approx 80~$keV), respectively. Note that 
Eq.~(\ref{E:anni_est}) neglects the increase in proton density due to
neutron decay.  The fraction $x$ is
well approximated by
$x\approx 1$,  since there are practically no protons present in the
annihilation region such that
most antibaryons annihilate on
neutrons. The results of the above estimate
agree remarkably well with the numerical results.

It is apparent from Eq.~\bref{E:anni_est} that the $n/p$-ratio at
$\The$ not only depends on the antimatter fraction $\RA$, but also on
the time when annihilation of the antimatter domains takes place. The
reason for this behavior is that the number of neutrons annihilated
is roughly independent of the annihilation time, but for early
annihilation this number is subtracted from a larger initial number
than in case of later annihilation.

The above estimate Eq.~\bref{E:anni_est} also predicts that it is
possible to completely avoid \he4, and in that case also 
\h2, \h3, $^3$He, and $^7$Li
synthesis, namely if $n/p|_{\The} =0$. (For antimatter fractions which
yield negative results for $n/p$ Eq.~\bref{E:anni_est} is obviously
not applicable.)  Thus there is no lower limit to the production of
\he4. An example of such a scenario is shown in panel (d) of
Fig.~\ref{F:np}. The antimatter fraction of $\RA=0.2$ exceeds the
neutron fraction at the time of annihilation ($\approx 0.5$~MeV) and
thus practically all neutrons are annihilated and no light element
nucleosynthesis is possible (cf. also the results for $\RA \gtrsim
0.2$ shown in Fig.~\ref{F:comb_2}).  This is to our knowledge the only
baryo-asymmetric scenario in which light-element nucleosynthesis is
absent~\cite{RJ:98}.

\subsection{Annihilation After \he4 Synthesis}
Essentially all free neutrons and antineutrons 
will be bound into \he4 and anti-\he4 nuclei at $\The$. At this time,
the neutron diffusion length is about 6~cm, thus antimatter domains
which are larger than $\approx$~6~cm may not annihilate induced by
mixing of matter and antimatter via neutron
diffusion. Further annihilation is delayed until transport of protons
and light nuclei over this distance is effective.  The proton
diffusion length does not grow to this value until the temperature
drops down to a few keV. But at this low temperature, the photon mean
free path has already increased enormously and thus baryonic density
gradients in the primordial fluid may not be supported any more by
opposing temperature gradients. Thus the regions far away from the
annihilation region, which are at high (anti-) baryon density,
quickly expand towards the baryon depleted and thus low-density
annihilation region and thereby transport matter and antimatter
towards the boundary. The annihilation time is thus controlled by the
hydrodynamic expansion time scale at late times.  Only the actual
mixing, i.e. the transport over the boundary, is still described by
baryon diffusion.

During the course of late time annihilation not only nucleon-antinucleon,
but also antinucleon-nucleus annihilations may take place. The elemental
abundances produced at the BBN epoch may now be substantially modified
not only by direct annihilations, but also due to the effects of the
secondary nuclei produced in antinucleon-nucleus annihilations. 
In particular, annihilations on \he4 produce \h2, \h3, 
and $^3$He nuclei, which, since they are energetic, 
may fuse via non-thermal nuclear reactions to form
$^6$Li (cf. Section~\ref{S:nucl}).
Furthermore, the
photodisintegration of \he4 by energetic photons arising in the
annihilation process becomes possible. 
These effects are evident from Figs.~\ref{F:comb_1}
and~\ref{F:comb_2}. Whenever $\ra \gtrsim 6$ cm, the yields for \h2 and
\he3, as well as for \li6 show a strong increase. The abundance of
\li7 is not much affected by late time annihilation, since for $R_A$
not too large there is no
efficient production channel leading to this isotope, and destruction
via direct annihilation is insignificant. 
The slightly elevated value for \li7/H
compared to the SBBN result is only due to our initial conditions
which lead to a higher value of $\eta$ during the BBN epoch for late
annihilation of high antimatter fractions. (\li7/H is an increasing
function of $\eta$ for $\eta \gtrsim 3 \times 10^{-10}$.)

Since annihilation of antimatter in a scenario
where antimatter is distributed in well-defined domains is mainly
confined to the region close to the matter-antimatter boundary, one
may speculate that secondary nuclei which are produced inside the
annihilation region are also annihilated.  But this is not necessarily
the case. The secondary nuclei gain a kinetic energy of the order of a
few tens of MeV in the \he4 disruption process~\cite{BBB:88}.  The
fraction which may survive the annihilation depends on the distance
the nuclei may free-stream away from the boundary before they 
thermalize. This fraction is initially small but increases rapidly due to an
increase of the stopping
length for energetic nuclei with cosmic time. (Of course, nuclei which 
travel into the antimatter domains and thermalize there will always be
annihilated.)
In particular, according to Eq.~\bref{E:lstop_ch2}
an energetic $^3$He nucleus of $50\,$MeV produced via annihilation
at $T\approx 3\,$keV has a stopping length of $\sim 1.5\,$cm. This
should be compared to the typical domain size $r_A^{100}\approx
10\,$cm of domains annihilating at $T\approx 3\,$keV (cf. 
Fig.~\ref{F:scales}), illustrating 
that the synthesized $^3$He nuclei will predominately be annihilated 
subsequently. In contrast, when the annihilation occurs at $T\approx
60\,$keV, for domain size $r_A^{100}\approx 10^3\,$cm, the stopping
length has increased to $\approx 4\times 10^3\,$cm implying that most
annihilation
produced $^3$He survives. These trends are evident from Figs.~\ref{F:comb_1}
and~\ref{F:comb_2} where for the same $R_A$ one observes a rapid increase
of production of \h2, $^3$He, and $^6$Li with increasing antimatter 
domain size. The increase is enhanced also due to the additional
production of \h2, $^3$He, and $^6$Li via photodisintegration of
$^4$He which becomes possible at low temperatures.

\begin{figure}
 \centerline{\psfig{figure=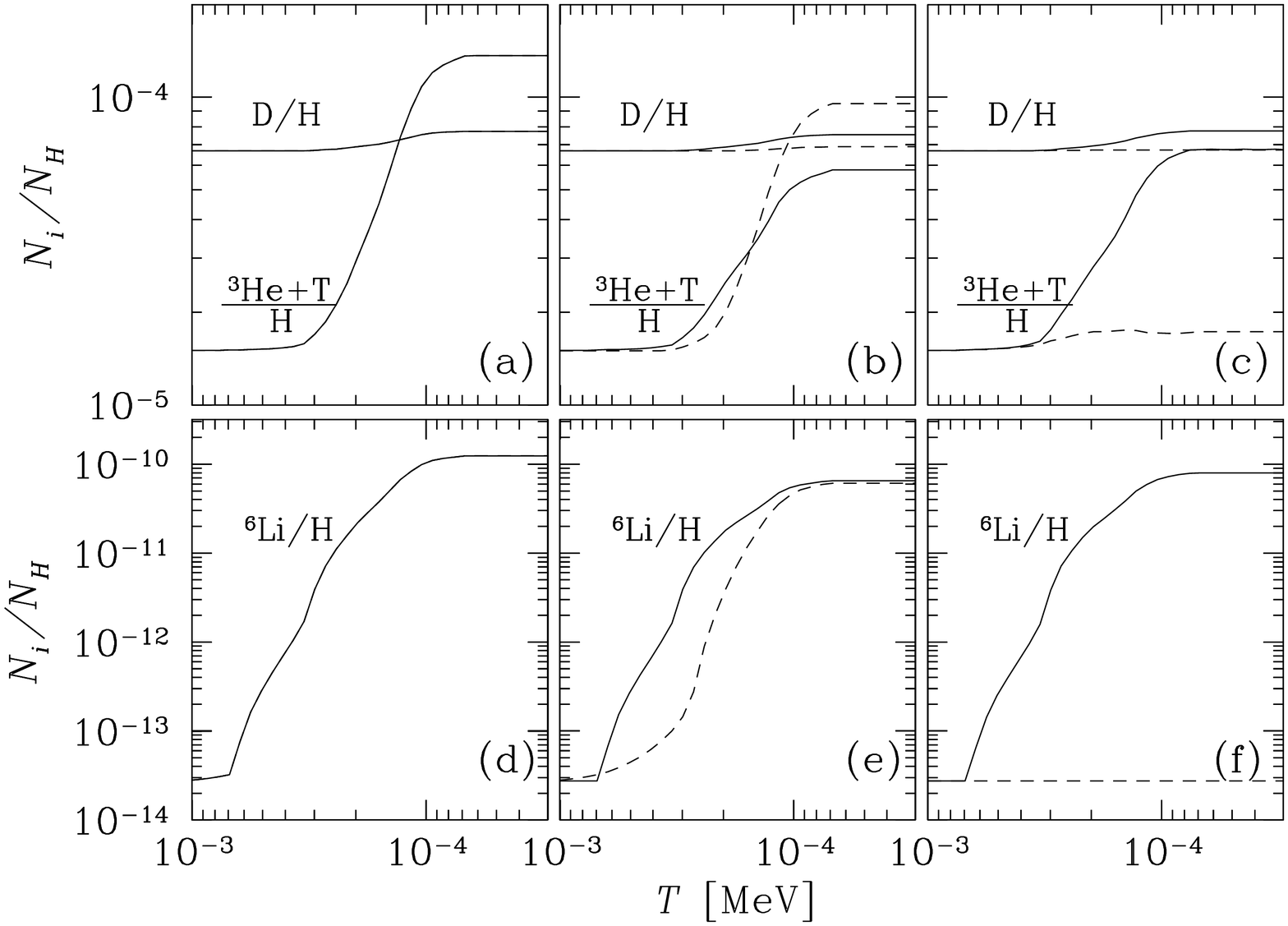,width=3.375in}}
{\caption{Abundance yields for \h2 and \he3 + \h3 (upper row), and
\li6 (lower row) obtained in consideration of different physical
effects. {\it Left column:} all effects included. {\it Middle column:}
only production by photodisintegration (dashed line) and only
direct production by annihilation and
escape from the annihilation region (solid line) taken into account.
{\it Right column:} only direct production by 
annihilation is taken into account, secondaries
remain either confined to the annihilation region (dashed line) or are
homogeneously distributed throughout the simulation volume (solid
line). See text for detailed discussion.
The antimatter fraction in the
simulations shown here is $\RA = 10^{-3}$ and the length scale $\ra =
5.5 \times 10^2$~cm.  \label{F:4comp}}}
\end{figure}

In order to gauge the relative importance of the two effects which
yield energetic \he3 nuclei, namely direct annihilation on \he4 and
photodisintegration of \he4, we show in Fig.~\ref{F:4comp} results for
simulations with $\RA= 10^{-3}$ and $\ra = 5.5 \times 10^2 $~cm, in
which annihilation occurs close to the temperature where the
photodisintegration yields are maximal ($T\approx 50$~eV). Shown are
the abundance yields for \h2 and \he3 (upper row) and for \li6 (lower
row).  In panels (a) and (d) all effects, production by
photodisintegration as well as direct production by annihilation and
escape from the annihilation region, are included. In the middle
column, panels (b) and (e), only one of the mechanisms is active at a
time. The solid line shows the results for \h2, \he3, and \li6 when
only direct production by annihilation is taken into account, while
the dashed line only considers \he4 photodisintegration induced
production of these elements.  We find that the photodisintegration
yields for \he3 are larger by nearly a factor of 2 than the direct
annihilation yields.  This is not surprising, given the peak
photodisintegration yield of about 0.1 \he3 nuclei per annihilation
(cf.~\cite{PSB:95}) and the probability of direct \he3 production in a
\pbar\he4 annihilation weighted by the relative abundance of \he4 to
protons; $ P_{\rm ^3He+^3H}\, ({^4{\rm He}}/p) \approx 0.05$
(cf. Tab.~\ref{T:probab}).  For the production of \li6 from energetic
\he3 and \h3 nuclei both effects are of the same importance, since the
yields for \li6 production via energetic nuclei generated by
photodisintegration of \he4 and annihilation of \he4 are similar over
a wide range of redshifts, $2 \times 10^4 \lesssim z \lesssim 4 \times
10^5$.  The remaining two panels, (c) and (f), demonstrate once more
the importance of the escape of the energetic secondaries from the
annihilation region. The dashed line shows the results of a simulation
where all annihilation generated nuclei are confined to the
annihilation region, while the solid line corresponds to a simulation
where the secondary nuclei are distributed homogeneously throughout
the simulation volume. Note that photodisintegration of \he4 was
ignored in these calculations. We may thus compare the solid lines in
panels (c) and (f) to the solid lines in panels (b) and (e). From
this, it is evident, that for annihilations occuring below $T\lesssim
100$~eV $\sim$ 80-90 \% of all secondary nuclei are able to escape from
the annihilation region and will thus survive subsequent annihilation.

\section{Discussion of the Results}\label{S:Res}
\subsection{Observational Constraints}
Any valid scenario for the evolution of the early Universe has to
reproduce the observationally inferred values of the light element
abundances, which we will summarize in the following.
The primordial \he4 mass fraction is commonly inferred from
observations of old, chemically unevolved
dwarf galaxies. Two distinct values are reported for the primordial
\he4 mass fraction $Y_p = 0.234 \pm 0.003^{\rm stat} \pm 0.005^{\rm
sys}$~\cite{OSS:97} and $Y_p = 0.244 \pm 0.002^{\rm stat}$
\cite{IT:98b}. Very recently two of the pioneers of the field have
determined the \he4 mass fraction in the Small Magellanic Cloud (SMC) by
observing 13 areas of the brightest H II region in that galaxy: NGC
346~\cite{PPR:00}. There observations are extrapolated to a value of 
$Y_p=0.2345\pm0.0026
(1\sigma)$. While in excellent agreement with the above quoted lower
value~\cite{OSS:97}, it is in conflict with the higher value
advertised in Ref.~\cite{IT:98b} and also not
compatible with the currently favored primordial \h2 determination.
There are three 
claimed detections of (\h2/H) ratios at high redshift from 
observations of quasar absorption line spectra.  
Similar to the case of \he4, two conflicting
values for the primordial (\h2/H) have been derived, $^2{\rm H/H} = 20 \pm
5 \times 10^{-5}$~\cite{We:97} and $^2{\rm H/H} = 3.39 \pm .25^{\rm
stat}_{2\sigma} \times 10^{-5}$~\cite{BT:98b}, with 
stronger observational support for the low \h2/H ratio. 
Using a new approach in
analyzing the spectra, Levshakov, Kegel and coworkers reported a
common value of $^2{\rm H/H} = 4.4\pm .3\times10^{-5}$~\cite{LKT:98a} for
all three absorption systems. 
Concerning the primordial $^3$He
abundance the situation is even less clear since only the chemically
relatively evolved $^3$He/H abundance in the
pre-solar nebula is available, and the chemical evolution of $^3$He
is poorly understood. It is, however,
reasonable to assume that the cosmic \he3/\h2 ratio is an
increasing function of time. Whenever \he3 is destroyed in stars, the more
fragile \h2 will certainly also be destroyed~\cite{SJSB:95}.
Thus $\left(\mhe3/^2{\rm H}\right)_t \gtrsim \left(\mhe3/^2{\rm H}\right)_p$ should hold for any time $t$ after 
Big Bang nucleosynthesis. Given the pre-solar \he3/\h2 ratio
\cite{Ge:93} we may impose the limit 
$\left(\mhe3/^2{\rm H}\right)_p\lesssim 1.$

Only traces of \li6 are produced in the framework of SBBN, and until
very recently the observational data was very sparse. For this reason,
\li6 was not considered to be a cosmological probe. With the 
confirmation of \li6 detections in old
halo-stars~\cite{SLN:93_HT:94_CSSVCA:99} and disk 
stars~\cite{NLPS:99} on the level $^6$Li/H $\sim 7\times 10^{12}$ this
may change.  
Nevertheless, since detection of such small $^6$Li/H abundances
requires operation close to the detection limits
of current instruments, and possible stellar depletion of \li6 is not
well understood, the use of $^6$Li as a cosmic probe may  be 
controversial at present. In this light, we adopt a tentative upper
bound on the primordial \li6 abundance of about $\mli6/{\rm H} \sim
7 \times 10^{-12}$. This limit may be used to
constrain some non-standard BBN scenarios, which greatly
overproduce \li6~\cite{Je:00}. 

\li7 is inferred at a remarkably constant abundance of A(\li7)$\equiv
\log_{10}(\mli7/{\rm H})-12 = 2.238\pm0.012\pm0.05^{\rm stat}$
\cite{BM:97} in old POP~II stars, referred to as the Spite-plateau.  
Nevertheless, stellar models which deplete \li7
considerably have been proposed~\cite{PDD:92,CD:94}. Recently, it has
been claimed that the primordial \li7 abundance should even be lower
than the plateau value~\cite{RNB:99,RBOFN:00}. As in the case of
\h2, the \li7 abundance is however not useful to derive limits
on our models. In the parameter range where the observationally
inferred \li7 abundance yields are violated, limits derived from other
elements are more stringent.
\subsection{Constraints on Antimatter Domains in the Early Universe}
\label{S:limits}
\begin{figure}
   \centerline{\psfig{figure=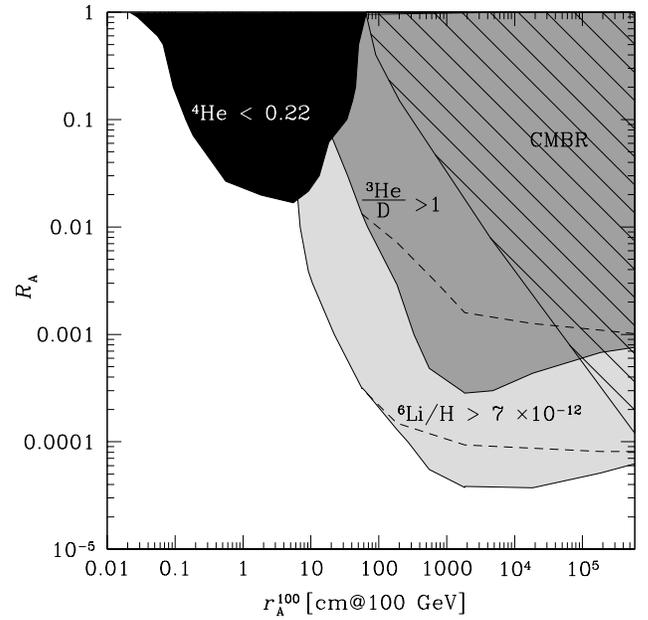,width=3.375in}}
{\caption{Limits on the presence of antimatter in the early
        Universe. Para\-meter combinations within the black shaded region
	result in a \he4 mass fraction below 0.22, while in the dark
	grey shaded region the bound ${\rm \mhe3/^2H} < 1$ is
	violated. The excluded range is extended by the light grey shaded
	region, if one adopts the tentative  bound ${\rm \mli6/H}
	< 7 \times 10^{-12}$. The dashed lines indicate the results
	when \he4 photodisintegration is ignored. Also shown are the
	usually weaker limits on the presence of antimatter from CMBR
	considerations (hatched region).
 \label{F:limits}}}
\end{figure}

In order to derive conservative limits on the amount of antimatter in
the early Universe, we first discuss our results with respect to
generally accepted observational constraints. While there is currently
a lively debate about which of the two independent \he4 determinations
reflects the primordial value, it seems reasonable to assume a \he4
mass fraction not lower than $Y_{p} \approx 0.22$~\cite{OSS:97,remark2}.  No
reliable limit on the \he3 abundance alone may be invoked, we
therefore use the constraint ${\rm \mhe3/^2H} < 1$~\cite{SJSB:95}. These
two values constitute our conservative data set as displayed
in Fig.~\ref{F:limits}. High antimatter fractions, $\RA \gtrsim 0.1$
may only be consistent with the observationally inferred light element
abundances if annihilation occurs close to weak freeze out, i.e. $\ra
\lesssim 10^{-1} $~cm. In this case, the weak interactions are still
rapid enough to at least partially reproduce the annihilated neutrons
and thus drive the $n/p$-ratio back towards the SBBN value. Antimatter
fractions larger than $\RA \gtrsim {\it a\, few}\, 10^{-2}$ on length
scales $\ra \gtrsim 10^{-1} $~cm result in an unacceptable low \he4
mass fraction, $Y_{p} < 0.22$, which is indicated by the black shaded
region in Fig.~\ref{F:limits}. Even larger antimatter regions, $\ra
\gtrsim 6 $~cm, annihilate at least partially via $\bar{p}$\he4
disruptions.  Since the destruction of only a minute fraction of \he4
leads to an observationally unacceptable enhancement of the \he3/\h2
ratio, the limits on the allowed antimatter fraction in this regime
may be as stringent as $\RA \lesssim {\it a\, few} \, 10^{-4}$ for
length scales $\ra \gtrsim 2 \times 10^2 $~cm (dark grey shaded
region in Fig.~\ref{F:limits}). 
Recently, Kurki-Suonio \& Sihvola~\cite{KS:99} derived similar
limits based on the  constraint  \he3/H~$\lesssim
10^{-4.5}$. We feel that this choice is not optimal,
given the large uncertainties in understanding the galacto-chemical
evolution of \he3.  Furthermore, they find a production of \he3  about a
factor of 3 higher than in our study.

If we employ the new and still slightly speculative \li6 limit
discussed above, we may significantly tighten the constraints on the
amount of antimatter by requiring that pre-galactic production of \li6
is not to exceed ${\rm \mli6/H} \lesssim 7 \times 10^{-12}$.  This
leads to an improvement of the limit on $\RA$ for late time
annihilation, i.e. $\ra \gtrsim 6$~cm, by up to two orders of
magnitude. This is evident from the light grey shaded region in
Fig.~\ref{F:limits}. Nevertheless, due to the loophole of possible
\li6 depletion in stars, and due to the still preliminary nature of
the \li6 observations, this limit should be regarded as tentative at
present.

The limits derived from annihilations below $T \lesssim 45$~eV,
corresponding to antimatter domain sizes of $\ra \gtrsim 10^3 \cm $,
have to be interpreted with care, since the photodisintegration yields
in that regime are uncertain due to the unknown photon spectrum, as we
discussed in Sec.~\ref{S:second_nucleon}. But even if we completely
ignore photodisintegration, meaningful limits due to direct production
of \he3 (and subsequent \li6 synthesis) via antiproton annihilation on
\he4 may still be obtained. These limits are indicated by the dashed
lines in Fig.~\ref{F:limits}.  Due to the increasing inefficiency of
photodisintegration at low temperatures, both limits converge for
large antimatter domain sizes.

Note that the limits derived here on the basis of the \he3/\h2 and \li6
data should be similar in magnitude to limits 
on scenarios where antimatter is
homogeneously injected into the plasma, for example by the decay of a
relic particle after the nucleosynthesis epoch, since they rely on the
production of secondary nuclei from \he4 disruption and
photodisintegration.  Both processes are generic for scenarios with
injection of antimatter. 
The competition of annihilation within, and escape from, the
annihilation region of the produced light isotopes is, however, particular to a
scenario with individual domains.  Escape of the annihilation products
is inefficient for domain sizes between $\sim
10 - 100\,$cm, corresponding to the \lq\lq injection\rq\rq\ of antimatter
between temperatures $\approx 3 - 0.4\,$~keV. In this regime more
stringent constraints would apply to a homogeneous injection of the 
antimatter. Furthermore, the reduction of the
$n/p$-ratio prior to \he4 synthesis and thus of the
\he4 mass fraction also only applies to models where antimatter is
confined to well defined domains. Only in this case annihilation
proceeds via baryon diffusion and thus the differential diffusion of
charged and neutral baryons may provide an efficient sink for
neutrons. In contrast, a homogeneous injection of antimatter
at temperatures above $T_{^4{\rm He}}\approx 80\,$keV (corresponding
approximately to the scale $\ra \approx 6\,$cm in Fig.~\ref{F:limits}) 
may be constrained by an increase of $Y_p$ due to proton-neutron
conversion induced by pion charge exchange~\cite{RS:88}.

For comparison, we have also shown in Fig.~\ref{F:limits} the limits
on annihilation which may be derived from the upper limits on
distortions of the spectrum of the CMBR. The very precise CMBR data
allows us to place constraints on the amount of non thermal energy
input at redshifts below $z \approx 3 \times 10^6$.  Each annihilation
transforms about one half of the rest mass of the particles into
electromagnetically interacting particles, 
thus the limits given in Ref.~\cite{FCGMSW:96}
may directly be converted into a limit on $\RA$, which is indicated by
the hatched region in Fig.~\ref{F:limits}.  Using the above
conservative data set, we find stronger limits from BBN than the ones provided
by the CMBR data for annihilations occuring at temperatures above
$T_{\rm ann} \gtrsim 1$~eV ($\ra \lesssim 10^5 $~cm),
corresponding to a redshift of $z \gtrsim 4 \times 10^3$. If we adopt
the new \li6 bound, the presence of antimatter is more tightly
constrained by BBN considerations, rather than by CMBR considerations,
for the whole parameter range down to the recombination epoch at
$z\approx 10^3$.
\subsection{Upper Limit on $\Omega_b$ in Matter-Antimatter
Cosmologies}\label{S:Omega}
It is of interest to contemplate if a BBN scenario with
matter-antimatter domains may reconcile the observationally inferred
element abundances with the theoretically predicted ones for a
baryonic density exceeding the upper bound from SBBN, $\Omega_b
h_{100}^2 \lesssim 0.02$. Possible alternative solutions to BBN which 
are in agreement with
observationally inferred abundances for higher values of $\Omega_b$ have
recently received renewed attention due to the results of the
{\sc Boomerang} and {\sc MAXIMA} experiments~\cite{Betal:00,BAetal:00} on the
anisotropies in the CMBR, 
which favor a baryonic
density exceeding the SBBN value~\cite{Letal:00_LP:00_TZ:00_Hetal:00}.
In the standard BBN scenario, such a
Universe suffers from overproduction of \he4 and \li7, and from severe
underproduction of \h2.  In a scenario with annihilating
antimatter domains, there exist two possibilities to reduce the
primordial \he4 mass fraction to the observed value. Early
annihilation, prior to \he4 synthesis, may reduce the $n/p$-ratio and
thus the final \he4 mass fraction. During late time
annihilation \he4 nuclei may be destroyed via antiproton induced
disruption and via photodisintegration.  At first sight, the
possibility to achieve observationally acceptable \he4 mass fractions
at high baryon-to-photon ratios looks promising. But upon closer
inspection, some severe shortcomings of such models arise.  Scenarios
at high net baryon-to-photon ratio and with annihilation prior to \he4
formation still overproduce \li7 relative to the observational
constraints. Furthermore, no additional source of \h2 exists in
this model, which is thus ruled out.  In the complementary case, where
annihilation is delayed until after the epoch of \he4 synthesis,
production of \h2 and \he3 due to disruption and
photodisintegration of \he4 results. Even though it is possible to find
models where late time \h2 production may reproduce the
observationally inferred value, the ratio of \he3/\h2 will exceed
unity. This is observationally unacceptable.  Further, such a scenario
would produce \li6 in abundance, which is most likely in conflict with
recent observations. This remains true, even if we drop the assumption
of a Universe in which the baryon- , or antibaryon-, to-photon ratio
has initially the same value throughout the Universe, and furthermore
allow the antimatter fraction and domain length scale to take
different values at different locations in space. Let us assume that
the Universe consists of two different types of regions. In regions of
type A with net baryon-to-photon ratio $\eta^{\rm A}_{\rm net}$, the
antimatter fraction is high, $\RA \lesssim 0.25$, and mixing is
effective between weak freeze out and $\The$. Irrespective of the
exact value for $\eta^{\rm A}_{\rm net}$, that region consists of
protons only after the annihilation is complete. In region B, with net
baryon-to-photon ratio $\eta^{\rm B}_{\rm net}$, antimatter domains
are larger, so they annihilate after the BBN epoch and light element
synthesis may take place.  If we further assume that region B is at
high baryonic density, $\eta^{\rm B}_{\rm net} \gg \eta_{\rm SBBN}$,
the production of \h2 and \he3 is negligible prior to
annihilation. Mass 2 and 3 elements will however be produced in
the course of late time annihilation of \pbar\ on \he4.  It is then
easily feasible to find a ratio between the volumes of the two regions
such that the average \he4 mass fraction $\bar{Y}_{\rm p}$ is diluted
to the observed value of $Y_p \approx 0.25$,
\begin{eqnarray}
\bar{Y}_{\rm p}&\approx&
\frac{\left(Y_p\right)_B}
{1+ \frac{\eta^{\rm A}_{\rm net} f_A}{\eta^{\rm B}_{\rm net}  f_B}}\,,
\end{eqnarray}
where $f_{A,B}$ are the fractions of space occupied by the two
different types of regions, respectively. The
\he4 mass fraction converges to $Y_p \approx 0.36$ for
high baryonic densities, the  required dilution factor is
thus at most $(Y_p)_B/Y_p \approx 1.5$ in order to
obtain $Y_p \approx 0.25$.  While the \he4 mass
fraction may agree with the observational constraints 
for an arbitrarily large
average baryon density, we face the same problems with production of
\h2, \he3 and \li6 via late time annihilation as discussed above
for a one zone model.  Furthermore, in region B, \li7 is produced well
in excess of the observed values and the dilution by mixing with the
proton-only zones of type A may not  reduce the \li7 abundance by more
than a factor of 1.5.
We thus conclude that it seems difficult to relax
the SBBN upper bound on $\Omega_b$ by the existence of
antimatter domains in the early Universe.

A further result of our study is that the putative presence of
antimatter in the early Universe may provide some relieve for the
tension between the lower of the two values for the primordial \he4
mass fraction, $Y_p = 0.234$~\cite{OSS:97}, and the low \h2
determination, $^2{\rm H/H} = 3.39 \times 10^{-5}$~\cite{BT:98b}. In
view of the recently reported value of $Y_p$ derived from observations
of the SMC~\cite{PPR:00}, which coincides with the low value, this
discrepancy has received new attention. In a Universe at a
comparatively high baryon-to-photon ratio of $\eta \approx 5 \times
10^{-10}$ with an antimatter
fraction of {\it a few} $10^{-4}$ distributed on length scales smaller
than 6~cm, the abundance yields for \he4 and \h2 may both be `low' and
thus the two observational constraints mentioned above may be
fulfilled simultaneously.

\section{Conclusions}\label{S:Disc}
In this work we have studied the mixing and subsequent annihilation of
antimatter domains in the early Universe during a period from a
cosmological temperature of about 10~MeV, well above the epoch of weak
freeze out, down to the formation of neutral hydrogen (recombination)
at about 0.2~eV.  Such distinct domains of antimatter may possibly arise in
some electroweak baryogenesis scenarios~\cite{CPR:94,GS:98ab}, as well as
in other proposed solutions to the baryogenesis problem (for a review 
see e.g.~\cite{Do:96}).
We have shown that the annihilation of antimatter domains may have
profound impact on the light element abundances.  Depending on the
time when annihilation occurs, we identify two main
effects. Annihilation prior to the incorporation of all neutrons into
\he4 results mainly in a reduction of the neutron-to-proton ratio,
which determines the amount of \he4 synthesized. Such scenarios are
thus constrained by the possible underproduction of \he4. Even more stringent
constraints on the antimatter-to-matter ratio may be derived if
antimatter
resides in slightly larger domains and
annihilation proceeds after the formation of \he4. In this case, the
dominant effect is the production of secondary energetic nuclei
(\h2, \h3, and \he3), which may  increase their
respective abundances, but may also lead to the production of \li6
nuclei. Further, energetic photons originating from the annihilation process
may produce additional energetic nuclei via the photodisintegration of
\he4. 

In a second aspect of our work we demonstrated that the presence of small
amounts of antimatter, separated from matter within some length scale
regime may, in fact, even improve the agreement between BBN theory and
observations by reducing the amount of synthesized $^4$He
while leaving other light isotope yields basically unaffected.
Finally, we argued that the SBBN upper bound on the cosmic baryon
density  $\Omega_{\rm b}$ is very unlikely to be relaxed in a scenario
with annihilating antimatter domains in the early Universe.

\acknowledgments
We want to thank K.~Protasov for helpful discussions and for providing
the annihilation data for the various \pbar$N$ reactions. 
This work was supported in part by the "Sonderforschungsbereich 375-95
f\"ur Astro-Teilchenphysik" der Deutschen Forschungsgemeinschaft.

\appendix
\section{\protect\\Structure of the Annihilation Region}\label{S:ann_regio}
In this appendix we will develop a detailed picture about the
structure of the annihilation region at the boundaries between matter and
antimatter domains. We will also illustrate why the numerical
resolution of this thin layer is not essential for making fairly
accurate prediction on the light-element nucleosynthesis in an environment with
matter-antimatter domains.

In case of annihilation via neutrons, i.e. $\Tanni
\gtrsim \The \approx 80 \keV$, diffusion within the annihilation
region is --- according to the local baryon-to-photon ratio
$\eta_{\rm ann}$ --- dominated either by magnetic moment scattering
on electrons and positrons ($\eta_{\rm ann} \lesssim 10^{-8}$), or by
nuclear scattering on protons ($\eta_{\rm ann} \gtrsim 10^{-8}$). In
both cases, the typical scattering time for neutrons is much smaller
than the annihilation time,
\begin{eqnarray}
\frac{\tau_{ne^\pm}}{\tau_{\rm ann}}&=&\frac
{\left(\sigma_{ne} v_b n_{e^\pm}\right)^{-1} }
{\left(\svanni\nbb^{\rm ann}\right)^{-1}}\nonumber  \\
&\approx&\frac
{\left(8 \times 10^{-4}  \mb \ \sqrt{T/m_{\rm N}}\right)^{-1}}
{\Big(40 \mb \ \eta_{\rm ann} \Big)^{-1}}\nonumber  \\&\approx&
6 \times 10^{-4}\left(\frac{T}{\MeV}\right)^{-1/2}
\left(\frac{\eta_{\rm ann}}{4 \times 10^{-10}}\right)
\label{E:ttne}\end{eqnarray}
and
\begin{eqnarray}
\frac{\tau_{np}}{\tau_{\rm ann}}&=&\frac
{\left(\sigma_{np} v_b n_p\right)^{-1} }
{\left(\svanni\nbb^{\rm ann}\right)^{-1}}\nonumber  \\
&\approx&\frac
{\left(2 \times 10^{4}  \mb \ \sqrt{T/m_{\rm N}}\right)^{-1}}
{\Big(40 \mb  \Big)^{-1}}\nonumber  \\&\approx&
6 \times 10^{-2}\left(\frac{T}{\MeV}\right)^{-1/2}\,.
\phantom{\left(\frac{\eta_{\rm ann}}{4 \times 10^{-10}}\right)}
\label{E:ttnp}
\end{eqnarray} 
Here $v_b$ is a typical baryon thermal velocity, $\nbb^{\rm ann}$
the antibaryon density in the annihilation region, $m_{\rm N}$  the
nucleon rest  mass and the relevant cross sections are
$\sigma_{ne}\approx 8 \times 10^{-4} \mb$ and 
$\sigma_{np}\approx 2 \times 10^{4} \mb$ (see e.g. Ref.~\cite{JF:94}).
Neutron scattering is
thus always more probable than annihilation.
Note that Eqs.~(\ref{E:ttne},~\ref{E:ttnp}) assume an
electron density  roughly equal to the photon
density, $n_{e^\pm} \approx \np$, which is appropriate at early times
i.~e. before $e^\pm$ annihilation, when neutron diffusion is important. 
In the numerical computations, however, we follow the exact densities of
the species.

Annihilation via induced by protons diffusion occurs 
only in the keV era, where proton
diffusion is limited by Thomson scattering of the electrons in the
 `electron-proton system' off the ambient photons. Even though
transport of the protons may now be controlled by hydrodynamic
expansion, the movement of the particles over the boundary and inside
the annihilation region is still described by diffusion. Comparing
the Thomson interaction time  with the
annihilation time for protons (cf. Eq.~\ref{E:cross_p_pbar}),
\begin{eqnarray}
\frac{\tau_{e\gamma}}{\tau_{\rm ann}}&=&\frac
{\left(\sigma_{e\gamma} v_b n_\gamma\right)^{-1} }
{\left(\svanni \nbb^{\rm ann}\right)^{-1}} \nonumber\\
&\approx&\frac
{\left(6.7 \times 10^{2}  \mb \ \sqrt{T/m_{\rm N}}\right)^{-1}}
{\Big(32 \mb \ \sqrt{\MeV/T} \, \eta_{\rm ann} \Big)^{-1}} \nonumber\\
& \approx&
6 \times 10^{-7}\left(\frac{T}{\keV}\right)^{-1}
\left(\frac{\eta_{\rm ann}}{4 \times 10^{-10}}\right),
\end{eqnarray} 
we find that the scattering time scale is again much shorter than the
annihilation time scale.
In both cases, the width of the annihilation region
$l^{\rm ann}$ is thus given by the distance
$d(\tau_{\rm ann})$ nucleons can
diffuse into the
respective anti-region during their typical lifetime against
annihilation $\tau_{\rm ann}$ (cf. Eq.~\ref{E:diff_len}),  
\begin{equation}
l^{\rm ann}\approx 2\,d(\tau_{\rm ann})
\approx 2\,\left(\int_0^{\tau_{\rm ann}} 6  D(t) \tot t 
\right)^{1/2} \approx 2 \sqrt{ 6
D \tau_{\rm ann}}\; . \label{E:l_anni}
\end{equation}
We have included a factor of 2 to allow for diffusion of matter into
the antimatter region as well as of antimatter into the matter region.
The diffusion constant $D$ can be taken to be constant over the
lifetime against annihilation.  In order to calculate $\tau_{\rm
ann}$, we need to estimate the density in the annihilation region.  We
assume that a steady state between diffusion of baryon number into the
annihilation region and annihilation of this baryon number is
established.  The concept of a steady state is only appropriate for
times somewhat shorter than the Hubble time, since the densities and
diffusion constants vary with the expansion of the Universe. A typical
baryonic density gradient some distance away from the annihilation
region will however always be of the order of $\Delta
n_b/d_b(\tau_{\rm Hubble})$, with $d_b(\tau_{\rm Hubble})$ the
diffusion length scale over one Hubble time.  The difference in baryon
density is given by $\Delta n_b = \tilde n_b- n_b^{\rm ann}$, with
$\tilde n_b$ the baryon density far away from the annihilation region
and $ n_b^{\rm ann}$ the baryon density within the annihilation
region.  The baryon density in the annihilation region will typically
be much smaller than $\tilde n_b$; therefore we replace $\Delta n_b$ by
$\tilde n_b$.  This leads us to approximate the baryon number flux
$F_b$ into the annihilation region by
\begin{eqnarray}
F_b&=& D \nabla  n_b A
\approx D  \frac{\Delta n_b}{d_b(\tau_{\rm Hubble})} A 
\approx D \frac{\tilde n_b}{d_b(\tau_{\rm Hubble})} A\; .
\end{eqnarray} 
The number of annihilations in a volume with surface
$A$ and width $l^{\rm ann}$ should then be  equal to  the flux
of baryons into the volume,
\begin{eqnarray}
\svanni  n_b^{\rm ann} \nbb^{\rm ann} \, A \,
l^{\rm ann} &=& 
D\frac{\tilde  n_b}{d_b(\tau_{\rm Hubble})} A\; .
\label{E:steady_st}
\end{eqnarray} 
As long as the diffusion length is considerably smaller than the size
of the antimatter region, $\tilde n_b$ is equal to the initial matter
density, $\tilde n_b=\bar\net \Delta^0 $, where $\Delta^0$ is the
initial baryon overdensity  and $\bar\net$ the
initial average net baryon density (see Eq.~\ref{E:net}).  We may
now compute the 
baryon density in the annihilation region. Inserting the
annihilation length $l^{\rm ann}$, Eq.~\bref{E:l_anni}, into
Eq.~\bref{E:steady_st} and using 
$\tau_{\rm ann} = (\svanni\nbb^{\rm ann})^{-1}$ yields
\begin{eqnarray}
\svanni  n_b^{\rm ann} \nbb^{\rm ann} 2 A 
\sqrt{\frac{6\, D}{\svanni\nbb^{\rm ann}}}
&=&
D\frac{\bar\net\Delta^0}{\sqrt{D\, \tau_{\rm Hubble}}} A\; .
\end{eqnarray} 
The baryon and antibaryon density within the annihilation region
should be of the same magnitude, thus we finally obtain for the baryon
density in the annihilation region 
\begin{eqnarray}
 n_b^{\rm ann}&=&\left(\frac{(\bar\net \Delta^0)^{2}}{6 \svanni
\tau_{\rm Hubble}}\right)^{1/3}.
\end{eqnarray} 
This may be written in terms of the local baryon overdensity
$\Delta^{\rm ann}$ in the annihilation region as 
\begin{eqnarray}
\Delta^{\rm ann}&\equiv& \frac{ n_b^{\rm ann}}{\bar\net}\nonumber\\&=&
2.4 \times 10^{-3}\, (\Delta^0)^{2/3}\left(\frac{ \svanni}{40\, \mb\,
c} \right)^{-1/3} \nonumber\\&&\times\left(\frac{\MeV}{T}\right)^{1/3}
\left(\frac{\eta}{4 \times 10^{-10}}\right).
\label{E:f_anni}
\end{eqnarray} 
Interestingly, the overdensity in the annihilation region
$\Delta^{\rm ann}$ is independent of the diffusion constant. We may
now calculate the width of the 
annihilation region in our comoving units,
\begin{eqnarray}
l^{\rm ann}_{100}&=&\frac{2}{R}\,\left(\frac{6\,D}
{\svanni \, \bar\net
\Delta^{\rm ann}}\right)^{1/2}.
\end{eqnarray}
Using the relevant diffusion constants  and 
annihilation cross sections  and further assuming
$\eta_{\rm net} = 4 \times 10^{-10}$, we obtain
\begin{eqnarray}
l^{\rm ann}_{100}&=&2 \times 10^{-4}\cm \,(\Delta^0)^{-1/3} 
\left(\frac{T}{\MeV}\right)^{-19/12}
\label{E:l_anni_n}
\end{eqnarray} 
for annihilation via neutron diffusion and 
\begin{eqnarray}
l^{\rm ann}_{100}&=&1.3 \times 10^{-1} \cm \, (\Delta^0)^{-1/3}
\left(\frac{T}{\keV}\right)^{-17/12}
\label{E:l_anni_p}
\end{eqnarray} 
for annihilation via proton diffusion. 
\begin{figure*}
 {\psfig{figure=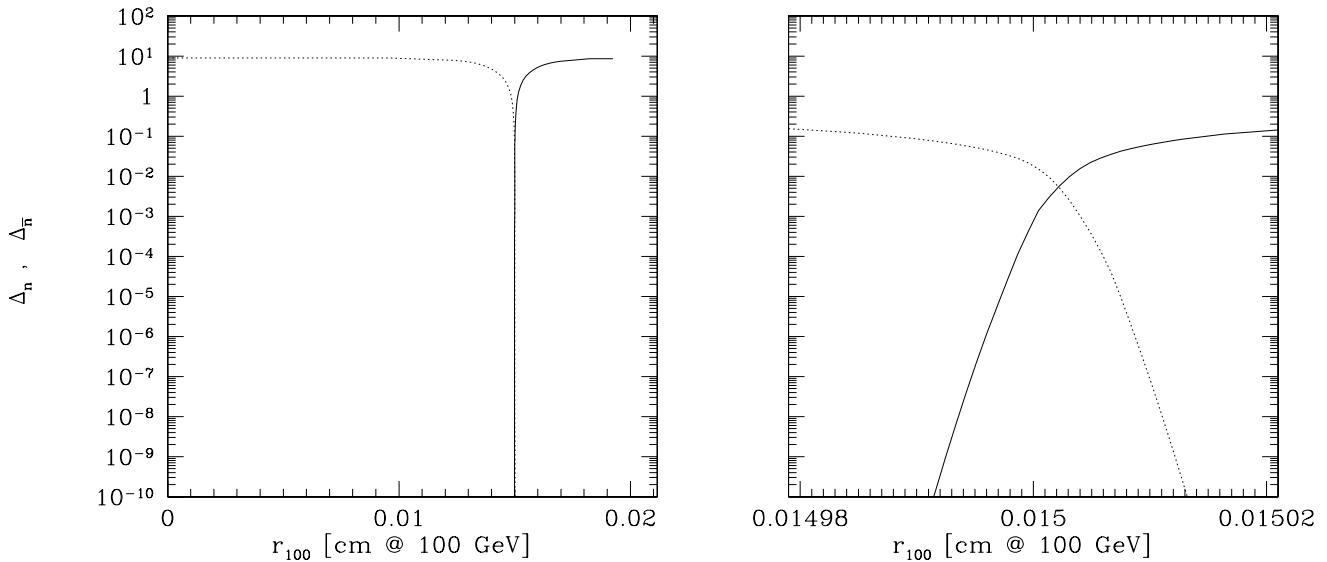,width=\textwidth}}
{\caption{{\it Left panel:} Snapshot of the neutron (full line) and
        the antineutron (dotted line) overdensity, $\Delta_{\rm n}$
        and $ \Delta_{\bar{n}}$ at a temperature of $T\approx 10 
	\MeV$. This distribution was obtained with a high resolution
	simulation; the antimatter parameters were $\RA$ = 0.9 and
	$\ra$ = $1.5 \times 10^{-2} \cm$.   {\it Right panel:} Zoom
	into the annihilation 	region. See text for discussion.
\label{F:l_anni_n}}}
\end{figure*}
We have numerically verified Eq.~\bref{E:l_anni_n} for a scenario with
antimatter regions of size $\ra = 1.5 \times 10^{-2} \cm$ and an
initial overdensity $\Delta^0\approx 10$, corresponding to a
matter-antimatter ratio of $\RA= 0.9$.  In order to check the validity
of the assumption of a steady state, we had to let the code evolve at
least over the period of one Hubble time.  The two snapshots of the
neutron and antineutron overdensity, $\Delta_{n}$ and $\Delta_{\bar
n}$ (cf. Fig.~\ref{F:l_anni_n}) were obtained in a simulation which
was started at $T\approx 20\MeV$, and evolved down to $T\approx
10\MeV$.  The left panel shows the whole simulation volume, while the
right panel is a zoom into the annihilation region of the same
simulation. The resolution is fine enough to describe (anti-)neutron
diffusion {\it within} the annihilation region. We find an overdensity
in the central region of $\Delta^{\rm ann}\approx \mbox{\it a\ few}\,
10^{-3}$ which may be compared to the above estimate, $\Delta^{\rm
ann}\approx 3 \times 10^{-3}$.  The width of the annihilation region
is $l^{\rm ann}_{100}\approx 10^{-6} \cm$, following
Eq.~\bref{E:l_anni_n}.
 
Since the two relevant processes  --- transport of particles through their
own region towards the annihilation region and diffusion within the
anti-region --- proceed on length scales which differ by orders of
magnitude, it is very time-consuming to run simulations with the
resolution necessary to adequately  describe both processes.
The numerical results presented in this work were obtained at a
resolution which properly resolves  the transport processes over the
distance of order of the domain size, but
does not resolve the diffusion within the annihilation region.  This
should however affect our results little, since the exact
composition of the annihilation region is not decisive for the final
abundances.

In case of annihilation before \he4 synthesis, the exact annihilation
time is crucial for our results.  Protons hardly play a role in case of early
annihilation due to their very short diffusion length. The protons which
are originally present in the annihilation region are quickly
annihilated.  Additional protons may not be transported into the
annihilation region and their density profile remains  frozen in.
The annihilation region is thus populated by neutrons and antineutrons
only, and further annihilation  may only  proceed via neutrons and
antineutrons.  All particles which reach the annihilation region will
inevitably be annihilated  on a very short time scale compared to the
transport time.  Thus the time scale for annihilation of all
antimatter is set by the transport of neutrons and antineutrons
towards the annihilation region, hence over considerably longer distances than
the annihilation region, which are properly resolved.

In case of annihilation after the disappearance of free neutrons at a
temperature of $\The \approx 80 \keV$, the dominant channels are
\pbar$p$ and \pbar\he4. The ratio of annihilations on either \he4 or
on protons is important, since this ratio determines how many
secondary nuclei, which arise in \he4 disruption, are produced for a
given antimatter fraction. This ratio depends again on the transport
of the nuclei over the whole matter region into the annihilation
region. The transport time scale may either be set by charged
particle diffusion or by hydrodynamic expansion. For both processes,
resolution of the whole simulation volume is important, but since again
all nuclei which reach the  annihilation region are inevitably
annihilated, the spatial distribution of the nuclei within the
annihilation region should be of negligible importance.

The effect of not resolving the annihilation region is that matter and
antimatter may travel further into the respective anti-region than is
physically correct. But since in both cases discussed above the number
of annihilations on a specific nucleus at a specific time is set by
the transport processes, this lack of resolution should not 
be relevant. The relative independence of the results on the exact
structure of the annihilation region is also evident by resolution
studies given in App.~\ref{S:num}.

Energetic secondary nuclei arising in the \he4 disruption process may
only escape from the annihilation region and thus influence the final abundance
yields if their stopping length is much larger than the annihilation
region. The correct treatment of this effect is therefore
independent of whether or not diffusion within the annihilation region
is resolved.

\section{Numerical Methods}\label{S:num}
The task of performing detailed numerical BBN calculations was first
fulfilled by Wagoner, Fowler \& Hoyle~\cite{WFH:67}. Jedamzik \&
Fuller~\cite{JF:94,JFM:94} developed an inhomogeneous BBN code to describe
the evolution of subhorizon-scale baryon-to-photon fluctuations in the
early Universe and the resultant modifications of the light element
abundances.  To this end, a Lagrangian grid of zones
was introduced in which the various thermodynamic quantities and the nuclear
densities may deviate from the respective horizon average value.  The
BBN network is coupled to all relevant hydrodynamic processes, such
as diffusion of baryons, photon diffusive heat transport, neutrino
heat transport and late-time hydrodynamic expansion of high-density
regions. The nuclear reaction network and the thermodynamic evolution
of the homogeneous radiation background is treated as in an updated
version of the original BBN code~\cite{Ka:92,SKM:93}.
Baryon diffusion and incorporation of baryons into nuclei proceeds on
fast time scales, it is thus necessary to treat baryon
diffusion implicitly. Further, neutrino and photon heat transport and
hydrodynamic processes are included in the code. 
\begin{figure}
   \centerline{\psfig{figure=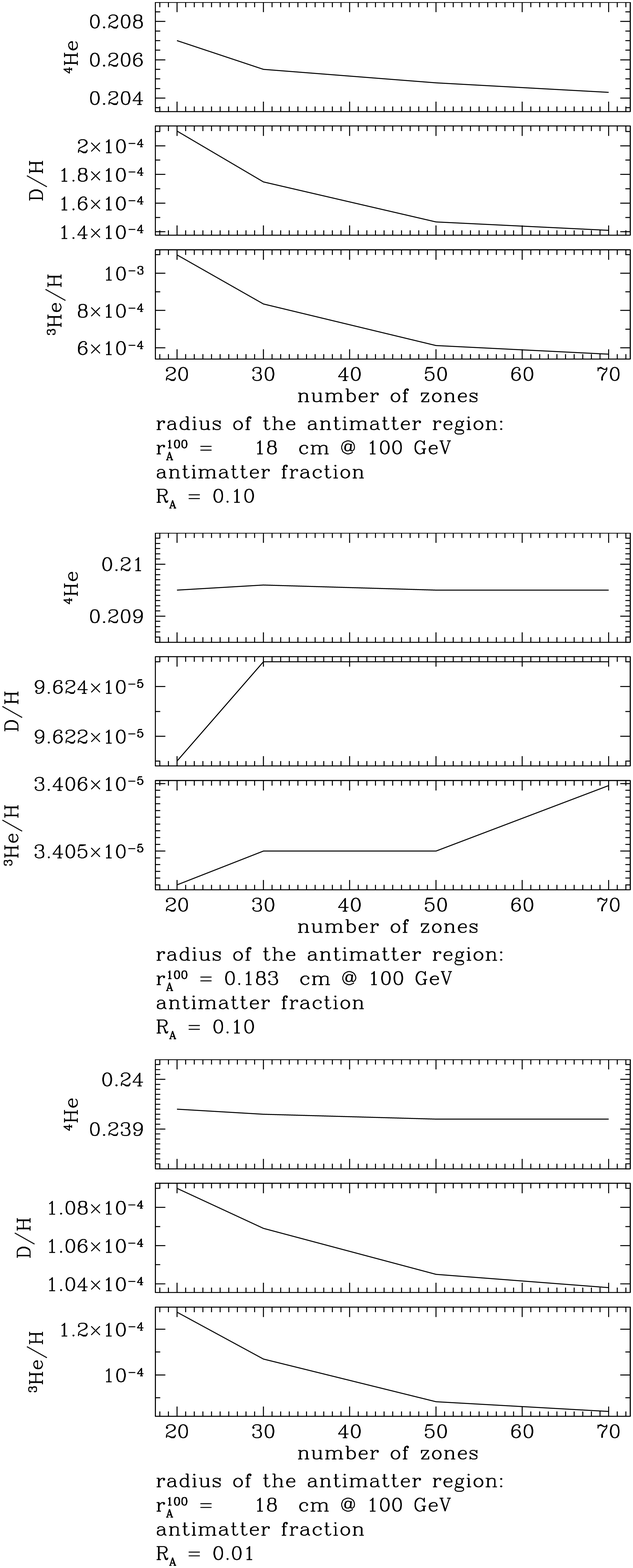,width=3.375in}}
{\caption{Light element
        abundances obtained in simulations with three different
        combinations of the antimatter parameters (see legend). The
        number of zones was varied between 20 and 70 to check for
        convergence of the results.
\label{F:con_zo}}}
\end{figure}

We extended the inhomogeneous code~\cite{JF:94,JFM:94} to
include antielements and adapted it to the present problem.
We use a set of concentric spherical shells to describe the
distribution of matter and antimatter. The
number of zones in the simulations has to be chosen such, that the
spatial resolution of the volume is sufficient to adequately describe
the relevant physical processes.  It turned out that 30 zones are
sufficient to resolve diffusion of the nuclei and obtain reasonable
convergence in the final abundances (see Fig.~\ref{F:con_zo}).  The
region where most of the annihilations occur, on the other hand, has
not been resolved in our simulations. The results should however be fairly
insensitive to the actual structure of the thin annihilation layer, as
we discuss in App.~\ref{S:ann_regio}.

The whole procedure of solving the nuclear reaction matrix and the
treatment of the hydrodynamical processes is included in a second order
Runge-Kutta scheme, i.e. it 
has to be done twice per time step. The results of the two Runge-Kutta
steps are averaged at the end of each time step. Since not only the
densities within each zone, but also the radii of the zones may
change, this has to be done very carefully in order to minimize errors.
An adequate independent check of the numerical simulations is the
achieved accuracy in baryon number conservation. Generally, the baryon
number 
is conserved on the level of $\Delta N_b/N_b \lesssim
{\cal O}(10^{-6})$ for those regions  of the parameter space relevant
for the derivation of our limits. Only for large antimatter fractions,
$\RA \gtrsim 0.5$ on length scales $\ra \gtrsim 10 \cm$, numerical conservation
of baryon number deteriorates. It seems however very unlikely to find
an observationally acceptable scenario for combinations of the
parameters in that range such that the simulation of such scenarios is
of little interest.

The escape of \li6 and the mass three nuclei from the annihilation
region was treated as follows. We keep track of the number of \h3, \he3
and \li6 nuclei produced in \pbar\he4 annihilations during a single
time step. Using the stopping length for these nuclei 
according to Eq.~\bref{E:lstop_ch2} we may calculate the fraction of those 
thermalized within the matter region by geometrical
considerations. This fraction is then added to
the number density of the respective nuclei.

\newcommand{\singleletter}[1]{#1}


\begin{thebibliography}{10}

\bibitem{St:76}
G. Steigman, Ann. Rev. Astron. Astroph. {\bf 14},  339  (1976).

\bibitem{CRG:98}
A. Cohen, A. de~R\'ujula, and S. Glashow, \apj {\bf 495},  539  (1998).

\bibitem{KKT:97}
W.~H. Kinney, E.~W. Kolb, and M.~S. Turner, \prl {\bf 79},  2620  (1997).

\bibitem{CR:98}
A.~G. {Cohen} and A. {de R\'ujula}, \apj {\bf 496},  L63  (1998).

\bibitem{DS:93}
A. Dolgov and J. Silk, \pr {\bf D47},  4244  (1993).

\bibitem{SZ:70b}
R.~A. Sunyaev and {\singleletter{Ya}}.~B. Zel'dovich, \apss {\bf 7},  20
  (1970).

\bibitem{CPR:94}
D. Comelli, M. Pietroni, and A. Riotto, Nuc. Phys. {\bf B412},  441  (1994).

\bibitem{GS:98ab}
M. Giovannini and M.~E. Shaposhnikov, \prl {\bf 80},  22  (1998).
M. Giovannini and M.~E. Shaposhnikov, Phys. Rev. {\bf D57},  2186  (1998).

\bibitem{KRS:00}
M. \singleletter{Yu}. Khlopov, S.~G. Rubin, and A.~S. Sakharov, hep-ph/0003285
  (2000).

\bibitem{RJ:98}
J.~B. Rehm and K. Jedamzik, \prl {\bf 81},  3307  (1998).

\bibitem{KS:99}
H. Kurki-Suonio and E. Sihvola, \prl {\bf 84},  3756  (2000).

\bibitem{St:72_CFL:75_Al:78}
G. Steigman, Cal-Tech Orange Aid {\bf OAP 280},    (1972);
F. {Combes}, O. {Fassi-Fehri}, and B. {Leroy}, \apss {\bf 37},  151  (1975);
J.~J. {Aly}, \aap {\bf 64},  273  (1978).

\bibitem{CKSZ:82_CKS:82}
V.~M. Chechetkin, M.~{\singleletter{Yu}}. Khlopov, M.~G. Sapozhnikov, and
  {\singleletter{Ya}}.~B. Zel'dovich, \pl {\bf B118},  329  (1982);
V.~M. Chechetkin, M.~{\singleletter{Yu}}. Khlopov, and M.~G. Sapozhnikov, Riv.
  Nuovo Cim. {\bf 5},  1  (1982).

\bibitem{BFPS:84}
{\singleletter{Yu}}.~A. Batusov {\it et~al.}, Nuovo Cim. Lett. {\bf 41},  223
  (1984).

\bibitem{KL:84_ENS:85_Li:86_Do:87_Ha:87}
M.~{\singleletter{Yu}}. Khlopov and A.~D. Linde, \pl {\bf B138},  265  (1984);
J. Ellis, D.~V. Nanopoulos, and S. Sarkar, Nuc. Phys. {\bf B259},  175  (1985);
D. Lindley, Phys. Lett. {\bf B171},  235  (1986);
R. {Do\-m\'{\i}n\-guez-Tenreiro}, \apj {\bf 313},  523  (1987);
I. Halm, Phys. Lett. {\bf B188},  403  (1987).

\bibitem{DY:87_YD:88}
R. {Do\-m\'{\i}n\-guez-Tenreiro} and G. {Yepes}, \apj {\bf 317},  L1  (1987);
G. {Yepes} and R. {Do\-m\'{\i}n\-guez-Tenreiro}, \apj {\bf 335},  3  (1988).

\bibitem{ZSKC:77_VDN:78}
{\singleletter{Yu}}.~B. {Zel'dovich}, A.~A. {Starobinskii},
  M.~{\singleletter{Yu}}. {Khlopov}, and V.~M. {Chechetkin}, Pis'ma Astron. Zh.
  {\bf 3},  208  (1977);
B.~V. {Vainer}, O.~V. {Dryzhakova}, and P.~D. {Naselskii}, Pis'ma Astron. Zh.
  {\bf 4},  344  (1978).

\bibitem{WFH:67}
R.~V. Wagoner, W.~A. Fowler, and F. Hoyle, \apj {\bf 148},  3  (1967).

\bibitem{SKM:93}
M.~S. Smith, L.~H. Kawano, and R.~A. Malaney, \apj\ Suppl. Ser. {\bf 85},  219
  (1993).

\bibitem{JF:94}
K. Jedamzik and G. Fuller, \apj {\bf 423},  33  (1994).

\bibitem{AHS:87}
J.~H. Applegate, C.~J. Hogan, and R.~J. Scherrer, \pr {\bf D35},  1151  (1987).

\bibitem{remark}
Here we have corrected for an obvious mistake in Ref.~\cite{JF:94}.

\bibitem{ADFMM:90}
C.~R. Alcock, D.~S. Dearborn, G.~M. Fuller, G.~J. Mathews, and B.~S. Meyer,
  \prl {\bf 64},  2607  (1990).

\bibitem{Pe:71}
P.~J.~E. Peebles, {\em Physical Cosmology} (Princeton University Press,
  Princeton, NJ, 1971).

\bibitem{BBB:88}
F. Balestra, S. Bossolasco, M.~P. Bussa, L. Busso, and L. Fava, Nuovo Cim. {\bf
  100A},  323  (1988).

\bibitem{CPZ:97}
J. Carbonell, K.~V. Protasov, and A. Zenoni, \pl {\bf B397},  345  (1997).

\bibitem{PBRZ:00}
K.~V. Protasov, G. Bonomi, E.~L. Rizzini, and A. Zenoni, Eur. Phys. J. {\bf
  A7},  429  (2000).

\bibitem{Zetal:00ab}
A. Zenoni {\it et~al.}, Phys. Lett. {\bf B461},  413  (1999); 
A. Zenoni {\it et~al.}, Phys. Lett. {\bf B461},  405  (1999).

\bibitem{MCK:88}
G.~S. Mutchler {\it et~al.}, \pr {\bf D38},  742  (1988).

\bibitem{CP:93}
J. Carbonell and K. Protasov, Hyperfine Interactions {\bf 76},  327  (1993).

\bibitem{CP:96}
J. Carbonell and K.~V. Protasov, Z. Phys. A {\bf 355},  87  (1996).

\bibitem{RS:88}
M.~H. Reno and D. Seckel, \pr {\bf D37},  3441  (1988).

\bibitem{Li:80}
D. Lindley, \mnras {\bf 193},  593  (1980).

\bibitem{EGLNS:92}
J. Ellis, G. Gelmini, J. Lopez, D. Nanopoulos, and S. Sarkar, Nuc. Phys. {\bf
  B373},  399  (1992).

\bibitem{PSB:95}
R. Protheroe, T. Stanev, and V. Berezinsky, \pr {\bf D51},  4134  (1995).

\bibitem{Je:00}
K. Jedamzik, \prl {\bf 84},  3248  (2000).

\bibitem{Ja:75}
J.~D. Jackson, {\em Classical Electrodynamics} (John Wiley \& Sons, New York,
  1975).

\bibitem{DEHS:88}
S. Dimopoulos, R. Esmailzadeh, L. Hall, and G. Starkmann, \apj {\bf 330},  545
  (1988).

\bibitem{remarkLi} The production of $^7$Li via  T + \he4 
$\to$ $^7$Li is less efficient due to a smaller cross section of
$\sim 0.1\,$mb. 


\bibitem{Me:72}
J.~P. Meyer, \aap\ Suppl. {\bf 7},  417  (1972).

\bibitem{JFM:94}
K. Jedamzik, G.~M. Fuller, and G.~J. Mathews, \apj {\bf 423},  50  (1994).

\bibitem{OSS:97}
K.~A. Olive, E. Skillman, and G. Steigman, \apj {\bf 483},  788  (1997).

\bibitem{IT:98b}
Y.~I. Izotov and T.~X. Thuan, \apj {\bf 500},  188  (1998).

\bibitem{PPR:00}
M. Peimbert, A. Peimbert, and M.~T. Ruiz, astro-ph/0003154  (2000).

\bibitem{We:97}
J.~K. Webb {\it et~al.}, \nat {\bf 388},  250  (1997).

\bibitem{BT:98b}
S. Burles and D. Tytler, \apj {\bf 507},  732  (1998).

\bibitem{LKT:98a}
S.~A. Levshakov, W.~H. Kegel, and F. Takahara, \apj {\bf 499},  L1  (1998).

\bibitem{SJSB:95}
G. Sigl, K. Jedamzik, D.~N. Schramm, and V.~S. Berezinsky, \pr {\bf D52},  6682
   (1995).

\bibitem{Ge:93}
J. Geiss,  in {\em Origin and Evolution of the Elements}, edited by N.
  Prantzos, E. Vangioni-Flam, and M. Cass\'e (Cambridge University Press,
  Cambridge, UK, 1993).

\bibitem{SLN:93_HT:94_CSSVCA:99}
V.~V. Smith, D.~L. Lambert, and P.~E. Nissen, \apj {\bf 408},  262  (1993);
L.~M. {Hobbs} and J.~A. {Thorburn}, \apj {\bf 428},  L25  (1994);
R. {Cayrel}, M. {Spite}, F. {Spite}, E. {Vangioni-Flam}, M. {Cass\'e}, and J.
  {Audouze}, \aap {\bf 343},  923  (1999).

\bibitem{NLPS:99}
P.~E. {Nissen}, D.~L. {Lambert}, F. {Primas}, and V.~V. {Smith}, \aap {\bf
  348},  211  (1999).

\bibitem{BM:97}
P. Bonifacio and P. Molaro, \mnras {\bf 285},  847  (1997).

\bibitem{PDD:92}
M.~H. Pinsonneault, C.~P. Deliyannis, and P. Demarque, \apj\ Suppl. Ser. {\bf
  78},  179  (1992).

\bibitem{CD:94}
B. {Chaboyer} and P. {Demarque}, \apj {\bf 433},  510  (1994).

\bibitem{RNB:99}
S.~G. {Ryan}, J.~E. {Norris}, and T.~C. {Beers}, \apj {\bf 523},  654  (1999).

\bibitem{RBOFN:00}
S.~G. {Ryan}, T.~C. {Beers},  K. A. {Olive}, B. D. {Fields}, and J.~E. {Norris}, \apj {\bf 530},  L57  (2000).

\bibitem{remark2} We note here that adopting the more conservative
constraint $Y_p > 0.20$, for example, would hardly change the
constraints on $R_A$ shown in Fig.~\ref{F:limits}. 

\bibitem{FCGMSW:96}
D.~J. Fixsen {\it et~al.}, \apj {\bf 473},  576  (1996).

\bibitem{Betal:00}
P. de~Bernardis {\it et~al.}, \nat {\bf 404},  955  (2000).

\bibitem{BAetal:00}
A. Balbi {\it et~al.}, astro-ph/0005124  (2000).

\bibitem{Letal:00_LP:00_TZ:00_Hetal:00}
A.~E. Lange {\it et~al.}, astro-ph/0005004  (2000);
J. Lesgourgues and M. Peloso, astro-ph/0004412  (2000);
M. Tegmark and M. Zaldarriaga, astro-ph/0004393  (2000);
S. Hanany {\it et~al.}, astro-ph/0005123  (2000).

\bibitem{Do:96}
A.~D. Dolgov,  in {\em Proceedings of International Workshop on Future
  Prospects of Baryon Instability Search in p-Decay and $n \rightarrow
  \overline{n}$ Oscillation Experiments, Oak Ridge, TN, USA ; 28 - 30 Mar
  1996}, edited by S.~J. Ball and Y.~A. Kamyshkov (National Laboratory, Oak
  Ridge, TN, 1996), p.\ 101.

\bibitem{Ka:92}
L. Kawano, FERMILAB-Pub-92/04-A  (1992).

\end{thebibliography}
\end{document}